\RequirePackage{fix-cm}
\documentclass[twocolumn, epjc3, comma, sort&compress, natbib]{svjour3}
\bibpunct{[}{]}{,}{n}{}{,} % to get "[numbered]" references from natbib

\journalname{Eur. Phys. J.}

\usepackage{latexsym}
\usepackage{amsmath}
\usepackage{amssymb}
\usepackage{amsfonts}
\usepackage{CJKutf8}

\usepackage[mathscr,scaled=1.15]{urwchancal}
\DeclareFontFamily{OT1}{pzc}{}
\DeclareFontShape{OT1}{pzc}{m}{it}%
{<-> s * [1.15] pzcmi7t}{}
\DeclareMathAlphabet{\mathpzc}{OT1}{pzc}{m}{it}

\usepackage{color}

\usepackage{supertabular}
\usepackage{placeins}
\usepackage{epsfig}
\usepackage{graphicx}

\definecolor{purple}{rgb}{0.5,0,0.5}
\definecolor{blue}{rgb}{0.0,0,0.9}
\definecolor{prdblue}{rgb}{0.133,0.118,0.498}
\usepackage[colorlinks=true, pdfstartview=FitV, linkcolor=prdblue, citecolor= prdblue, urlcolor=prdblue]{hyperref}

\makeatletter

\setbox0\hbox{$\xdef\scriptratio{\strip@pt\dimexpr
    \numexpr(\sf@size*65536)/\f@size sp}$}

\newcommand{\scriptveryshortarrow}[1][3pt]{{%
    \hbox{\rule[\scriptratio\dimexpr\fontdimen22\textfont2-.2pt\relax]
               {\scriptratio\dimexpr#1\relax}{\scriptratio\dimexpr.4pt\relax}}%
   \mkern-4mu\hbox{\let\f@size\sf@size\usefont{U}{lasy}{m}{n}\symbol{41}}}}

\makeatother

\begin{document}

\begin{CJK}{UTF8}{song}

%Title of paper
%\title{Symmetry-preserving calculation of the pion's valence-quark distribution}
%\title{Unification of continuum and lattice predictions of the pion's valence distribution}
%\title{Symmetry, symmetry breaking, and pion parton distributions}
\title{$\,$\\[-6ex]\hspace*{\fill}{\normalsize{\sf\emph{Preprint no}.\ NJU-INP 053/21}}\\[1ex]
%E615}
Concerning pion parton distributions}

\author{Z.-F.~Cui\thanksref{NJU,INP}
       \and
       M.~Ding\thanksref{HZDR}
        \and
        J.\,M.~Morgado\thanksref{UHue}
        \and
        K.~Raya\thanksref{UGr,UNAM}
        \and
       D.~Binosi\thanksref{ECT}
       \and
       L.~Chang\thanksref{NKU}
       \and
       J.~Papavassiliou\thanksref{UVal}
       \and
       C.\,D.~Roberts\thanksref{NJU,INP}
       \and
       J.~Rodr\'{\i}guez-Quintero\thanksref{UHue}
       \and
       S.\,M.~Schmidt\thanksref{HZDR,RWTH}
}

%\thankstext{eYL}{luya@nju.edu.cn}
%\thankstext{eDB}{binosi@ectstar.eu}
%\thankstext{eMD}{mding@ectstar.eu}
%\thankstext{eCDR}{cdroberts@nju.edu.cn}
%\thankstext{eHYX}{hyxing@smail.nju.edu.cn}
%\thankstext{eCX}{cxu@nju.edu.cn}

\authorrunning{Z.-F.~Cui \emph{et al}.} % if too long for running head

\institute{School of Physics, Nanjing University, Nanjing, Jiangsu 210093, China \label{NJU}
           \and
           Institute for Nonperturbative Physics, Nanjing University, Nanjing, Jiangsu 210093, China \label{INP}
           \and
           Helmholtz-Zentrum Dresden-Rossendorf, Dresden D-01314, Germany \label{HZDR}
           \and
           Department of Integrated Sciences and Center for Advanced Studies in Physics, Mathematics and Computation, \\
           \hspace*{0.5em}University of Huelva, E-21071 Huelva; Spain \label{UHue}
           \and
           Departamento de F\'isica Te\'orica y del Cosmos, Universidad de Granada, E-18071, Granada, Spain \label{UGr}
           \and
           Instituto de Ciencias Nucleares, Universidad Nacional Aut\'onoma de M\'exico, Apartado Postal 70-543, CDMX 04510, M\'exico \label{UNAM}
           \and
           European Centre for Theoretical Studies in Nuclear Physics and Related Areas, \\ \mbox{$\;\;$}Villa Tambosi, Strada delle Tabarelle 286, I-38123 Villazzano (TN), Italy \label{ECT}
           \and
           School of Physics, Nankai University, Tianjin 300071, China \label{NKU}
           \and
           Department of Theoretical Physics and IFIC, University of Valencia and CSIC, E-46100, Valencia, Spain \label{UVal}
           \and
           RWTH Aachen University, III. Physikalisches Institut B, Aachen D-52074, Germany \label{RWTH}
\\[1ex]
Email:
%\email[]{phycui@nju.edu.cn}
%\email[]{m.ding@hzdr.de}
%\email[]{josemanuel.morgado@dci.uhu.es}
%\email[]{khepani@ugr.es}
\href{mailto:binosi@ectstar.eu}{binosi@ectstar.eu} (D.~Binosi);
\href{mailto:leichang@nankai.edu.cn}{leichang@nankai.edu.cn} (L.~Chang);
\href{mailto:Joannis.Papavassiliou@uv.es}{Joannis.Papavassiliou@uv.es} (J.~Papavassiliou);
\hspace*{3.1em}\href{mailto:cdroberts@nju.edu.cn}{cdroberts@nju.edu.cn} (C.\,D.~Roberts);
\href{mailto:jose.rodriguez@dfaie.uhu.es}{jose.rodriguez@dfaie.uhu.es} (J.~Rodr\'{\i}guez-Quintero);\\
\hspace*{3.1em}\href{mailto:s.schmidt@hzdr.de}{s.schmidt@hzdr.de} (S.\,M.~Schmidt)
            }

%Collaboration name if desired (requires use of superscriptaddress
%option in \documentclass). \noaffiliation is required (may also be
%used with the \author command).
%\collaboration can be followed by \email, \homepage, \thanks as well.
%\collaboration{}
%\noaffiliation

\date{2021 December 16}
%\date{2021 November 30}
%\date{2021 November 20}
%\date{2021 November 12}

\maketitle

\end{CJK}

\begin{abstract}
Analyses of the pion valence-quark distribution function (DF), ${\mathpzc u}^\pi(x;\zeta)$, which explicitly incorporate the behaviour of the pion wave function prescribed by quantum chromodynamics (QCD), predict ${\mathpzc u}^\pi(x\simeq 1;\zeta) \sim (1-x)^{\beta(\zeta)}$, $\beta(\zeta \gtrsim m_p)>2$, where $m_p$ is the proton mass.  Nevertheless, more than forty years after the first experiment to collect data suitable for extracting the $x\simeq 1$ behaviour of ${\mathpzc u}^\pi$, the empirical status remains uncertain because some methods used to fit existing data return a result for ${\mathpzc u}^\pi$ that violates this constraint.  Such disagreement entails one of the following conclusions: the analysis concerned is incomplete; not all data being considered are a true expression of qualities intrinsic to the pion; or QCD, as it is currently understood, is not the theory of strong interactions.  New, precise data are necessary before a final conclusion is possible.  In developing these positions, we exploit a single proposition, \emph{viz}.\ there is an effective charge which defines an evolution scheme for parton DFs that is all-orders exact.  This proposition has numerous corollaries, which can be used to test the character of any DF, whether fitted or calculated.
%
%\rule{0.78\linewidth}{0.1ex}
\end{abstract}
%%
%%Keywords:
%%proton \sep
%%magnetic charge radius \sep
%%electric charge radius \sep
%%emergence of mass \sep
%%lepton-hadron scattering \sep
%%strong interactions in the standard model of particle physics

%%%%%%%%%%%%%%%%%%%%%%%%%%%%%%%%%%%%%%%%%%%%%%%%%%%%%%%%%%%%%%%%%%%%%%%%%%%%%%%%%%%%%%%%%%%%%%%%%%%%%%%%%%%%%%%%%%%%%%%
% 4500 words

%\noindent\emph{1.$\;$Introduction} ---
\section{Introduction}
\begin{figure}[b]
\centerline{\includegraphics[width=0.27\textwidth]{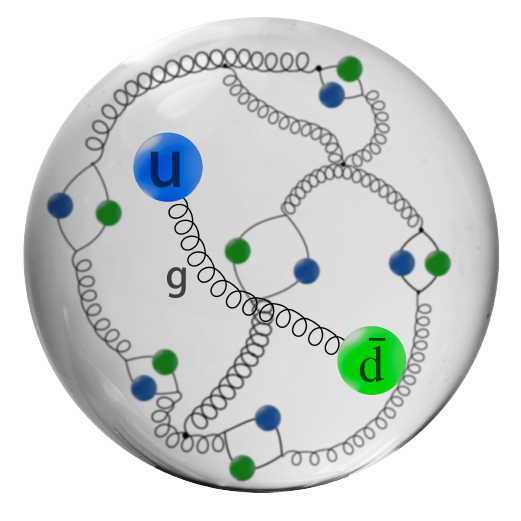}}
\caption{\label{ImagePion}
Referring to QCD's Lagrangian degrees-of-freedom, the $\pi^+$ is built from one valence $u$-quark, one valence $\bar d$-quark, and, owing to the properties of QCD, infinitely many gluons and sea quarks, drawn here as ``springs'' and closed loops, respectively.
The $\pi^-$ is $d\bar u$ and the $\pi^0$ is $u \bar u - d\bar d$.
}
\end{figure}

Amongst all the hadrons that appear in Nature's catalogue of bound states, pions would seem to be the simplest.  There are three such states ($\pi^\pm$, $\pi^0$) and within the Standard Model (SM) they are described as systems built from a valence-quark and a valence antiquark, \emph{e.g}., as illustrated in Fig.\,\ref{ImagePion}, the positively charged pion, $\pi^+$, is a $u \bar d$ state: one valence $u$-quark and one valence $\bar d$-quark.  There are complications, however.  In the absence of Higgs boson couplings into quantum chromodynamics (QCD) -- the SM's strong interaction part, pions are massless Nambu-Goldstone bosons.  Even with Higgs couplings restored, $u$, $d$ quarks remain light; and so do pions \cite{Zyla:2020zbs}, possessing masses, $m_{\pi^\pm}\approx m_{\pi^0}$, far less than that of the proton, $m_p$.  Amongst other things, these features are critical to the stability of all nuclei \cite{Yukawa:1935xg} and, therefore, to the emergence of the known Universe.

Indeed, Nature has no more fundamental Nambu-Goldstone boson than the pion.  If it is a QCD bound state, then it must have a complex structure -- Fig.\,\ref{ImagePion}, because although QCD has not been solved, many remarkable properties have been exposed.
For instance, gluons, which appear as massless gauge-boson degrees-of-freedom in the QCD Lagrangian, acquire a momentum dependent mass function, $m_g(k^2)$, whose value on $k^2\simeq 0$ is characterised by a renormalisation group invariant mass $m_0 \approx m_p/2$ \cite{Cornwall:1981zr, Aguilar:2008xm, Binosi:2016nme, Gao:2017uox, Fischer:2018sdj, Binosi:2019ecz, Cui:2019dwv, Roberts:2021nhw}.  This is a primary sign of the dynamical violation of scale invariance in QCD \cite{Roberts:2016vyn}, whose origin is strong gluon self-interactions, and the consequent emergence of hadron mass (EHM); namely, the appearance of a nuclear-size mass-scale in a theory that, \emph{a priori}, contains no masses at all.

In connection with the Nambu-Goldstone pions, two key corollaries of EHM are the appearance of a mass-function for quarks, whose value on $k^2\simeq 0$ is roughly $m_p/3$ \cite{Binosi:2016wcx}, and the intimate relationship \cite{Lane:1974he, Politzer:1976tv, Maris:1997tm, Qin:2014vya, Williams:2015cvx, Binosi:2016rxz} between that mass function and the pion's Poincar\'e-co\-va\-riant bound-state wave function, $\chi_\pi(k,k-P)$, where $P$ is the pion's total momentum and $k$ is the momentum of the valence-quark.
Combined, these two properties entail that although it can be exposed using a wide variety of means \cite{Carman:2020qmb, Brodsky:2020vco, Barabanov:2020jvn, Roberts:2020hiw}, the cleanest observable manifestations of EHM are to be found in pion properties; so imparting utmost importance to the goals of revealing and explaining pion structure \cite{JlabTDIS1, Adams:2018pwt, Aguilar:2019teb, Chen:2020ijn, Anderle:2021wcy, Arrington:2021biu}.
In this connection, much can be learnt from data that may be used to extract pion parton distribution functions (DFs).

Section~\ref{SecLargeX} explains QCD constraints on the behaviour of the pion valence-quark DF at large values of the light-front longitudinal momentum fraction, $x$; and Secs.\,\ref{SecExpt}, \ref{SecFits} review aspects of existing experiments \cite{Corden:1980xf, Badier:1983mj, Betev:1985pg, Conway:1989fs} and associated phenomenological fits \cite{Aicher:2010cb, Novikov:2020snp, Han:2020vjp, Barry:2021osv} which relate to the determination of this behaviour.
Section~\ref{SecDYdata} presents a novel perspective on the data in Ref.\,\cite[E615]{Conway:1989fs}, referring it to the character of the quark+antiquark interaction; and also describes the importance and expression of key QCD global symmetries in calculations of ${\mathpzc u}^\pi$, along with a natural definition for the hadron scale, $\zeta_{\cal H}$.
The principle and practical implications of all-orders evolution are detailed in Sec.\,\ref{SecAOE}, enabling an assessment of contemporary fits to data and the methods they employ in Secs.\,\ref{NLLdM}, \ref{NLLMF}.
Section~\ref{epilogue} provides a summary and an outlook.

%\medskip

%\noindent\emph{2.$\;$Pion valence-quark distribution at large $x$} ---
\section{Pion valence-quark distribution at large $\mathbf x$}
\label{SecLargeX}
In quantum mechanics, a bound-state's Schr\"odinger\linebreak  wave function contains a great deal of information about the system.  Owing to the complexities of quantum field theory, the nearest analogue for the pion is its light-front wave function (LFWF) \cite{Coester:1992cg, Brodsky:1997de}, $\psi_\pi(x,\vec{k}_\perp;P)$,\linebreak which can be obtained via a light-front projection of $\chi_\pi(k,k-P)$ \cite{tHooft:1974pnl, Chang:2013pq} and permits interpretation as a probability amplitude, unlike the Bethe-Salpeter wave function.  Here, using linearly independent light-like four-vectors $n$, $\bar n$, with $n^2=0=\bar n^2$, $n\cdot \bar n=-1$:
$x=n\cdot k/n\cdot P$, \emph{i.e}., the light-front fraction of the pion's total momentum carried by the valence-quark;
and
$k_\perp$ is the two-vector built from the in-general nonzero components of
$k_\mu = O^\perp_{\mu\nu}k_\nu$, $ O^\perp_{\mu\nu} = \delta_{\mu\nu} + n_\mu \bar n_\nu + \bar n_\mu n_\nu$, \emph{viz}.\ that part of the valence-quark's momentum which lies in the light-front transverse plane.
When discussing $\psi_\pi$, it is common, although not mandatory, to consider its expansion over an infinite dimensional Fock space of basis states built using non-interacting gluon and quark partons \cite{Brodsky:1989pv}.  Then the expansion coefficient associated with any particular configuration of gluon and quark partons is the probability amplitude for finding the hadron with the identified partonic configuration carrying and sharing all its properties.

A hadron's LFWF provides direct access to one of the first quantities used to reveal the presence of quarks within hadrons \cite{Taylor:1991ew, Kendall:1991np, Friedman:1991nq, Friedman:1991ip}; namely, the hadron's parton DFs \cite{Feynman:1969wa}.  As an example, ${\mathpzc u}^\pi(x;\zeta)$ describes the probability density for finding a valence $u$-quark with light-front momentum fraction $x$ when the pion is resolved at scale $\zeta$.  Notably, at any finite scale, this valence-quark will not be equivalent to a valence-quark-parton; instead, the valence-quark will be linked to the quark-parton as an object dressed by QCD interactions in the manner described by the quark gap equation \cite{Roberts:1994dr}.

Using contemporary tools:
\begin{equation}
\label{Eqvalence}
{\mathpzc u}^\pi(x;\zeta) \stackrel{x\in(0,1)}= H_\pi^{\mathpzc u}(x,t=0;\zeta)\,,
\end{equation}
where $H_\pi^{\mathpzc u}$ is the valence $u$-quark generalised parton distribution (GPD), which, on the kinematic domain of interest herein (zero skewness), may be expressed in terms of the pion's valence-quark LFWF \cite{Diehl:2000xz}:
\begin{align}
H_\pi^{\mathpzc u}(x,t;\zeta)
&= \int \frac{d^2{k_\perp}}{16 \pi^3}
\psi_{\pi}^{u\ast}\left(x,{k}_{\perp +}^2;\zeta \right)
%\nonumber \\ && \rule[0cm]{3.25cm}{0cm} \times
\psi_{\pi}^{u}\left( x,{k}_{\perp -}^2;\zeta \right) \,,
\label{eq:overlap}
\end{align}
where $k_{\perp \pm}=k_\perp \pm (1-x)t/2$, with $t$ the squared mo\-men\-tum-transfer to the pion in processes designed to measure the GPD.  In the ${\cal G}$-parity symmetry limit, which is an accurate reflection of Nature, $\bar {\mathpzc d}^\pi(x;\zeta)={\mathpzc u}^\pi(x;\zeta)$.

Since calculation of any bound-state wave function is a nonperturbative problem, perturbative QCD (pQCD) cannot be used to calculate the $x$-dependence of ${\mathpzc u}^\pi(x;\zeta) $.  However, it can be employed to predict the behaviour on $x\simeq 1$.   Owing to implicit kinematic constraints on this domain, only the leading two-body Fock-space component of the pion's complete LFWF is sampled, enabling simplifications that lead to the result
\begin{equation}
\label{pionDFpQCD}
{\mathpzc u}^\pi(x;\zeta) \stackrel{x\simeq 1}{\sim} (1-x)^{\beta \,=\,2+\gamma(\zeta)}\,,
\end{equation}
where $\gamma(\zeta) \geq 0$ grows logarithmically with $\zeta$, expressing the impact of gluon radiation from the struck quark that is realised in QCD evolution equations (DGLAP) \cite{Dokshitzer:1977sg, Gribov:1971zn, Lipatov:1974qm, Altarelli:1977zs}.
The hadron scale is that value of $\zeta=\zeta_{\cal H}$ at which begins gluon emission from the valence quarks in the complete wave function \cite{Brodsky:1979gy}: $\gamma(\zeta_{\cal H})=0$.

Eq.\,\eqref{pionDFpQCD} has been derived in various ways, \emph{e.g}.,\linebreak Refs.\,\cite{Ezawa:1974wm, Farrar:1975yb, Soper:1976jc, Brodsky:1979gy, Yuan:2003fs}; and as we now indicate, it can also be obtained straightforwardly from Eq.\,\eqref{eq:overlap}.  The value of the integral in Eq.\,\eqref{eq:overlap} is determined by the large-$k_\perp^2$ behaviour of $\psi_{\pi}^{u}\left( x,{k}_\perp^2;\zeta \right)$, which may be obtained using the operator product expansion \cite{Brodsky:1980ny}:
\begin{equation}
\label{OPE}
\psi_{\pi}^{u}\left( x,{k}_\perp^2;\zeta \right) \stackrel{k_\perp^2 \gg m_0^2}{=} d_0(\zeta) \, x(1-x)\frac{\alpha(k_\perp^2)}{k_\perp^2} + \ldots\,,
\end{equation}
where $ d_0(\zeta)$ measures the strength of the leading term at resolving scale $\zeta$, $\alpha(k_\perp^2)$ is the QCD running coupling, and the ellipsis indicates omitted terms that are suppressed by additional powers of $1/\ln(k_\perp^2/\Lambda^2_{\rm QCD})$, where $\Lambda_{\rm QCD}$ is a renormalisation group invariant mass scale whose size is an expression of EHM.  Inserting Eq.\,\eqref{OPE} into Eq.\,\eqref{eq:overlap} and recognising that no infrared divergence can occur because all (regularised or renormalised) Fock-space elements are independently normalisable, then Eq.\,\eqref{pionDFpQCD} is recovered via Eq.\,\eqref{Eqvalence}.

It is now evident that Eq.\,\eqref{pionDFpQCD} expresses intrinsic properties of the pion LFWF.  This fact is expressed in \emph{all} existing calculations, as highlighted elsewhere \cite{Holt:2010vj, Fanelli:2016aqc, deTeramond:2018ecg, Lan:2019rba, Chang:2014lva, Ding:2019qlr, Ding:2019lwe, Kock:2020frx, Chang:2020kjj, Zhang:2020ecj, Cui:2020dlm, Cui:2020tdf, Chang:2021utv, Zhang:2021tnr}.  Indeed, working in the chiral limit and assuming that chiral symmetry is dynamically broken, so $m_\pi^2=0$ in the presence of a dynamically generated dressed-quark mass \cite{Lane:1974he, Politzer:1976tv, Maris:1997tm, Qin:2014vya, Williams:2015cvx, Binosi:2016rxz}, one can show algebraically \cite[Sec.\,5A]{Roberts:2021nhw} that when quark+antiquark scattering in a four-dimensional theory is mediated by the exchange of a vector boson whose propagator is monotonically decreasing and behaves as $[1/k^2]^{n\geq 0}$ on $k^2 \gg m_p^2$, then\footnote{The cases $0\leq n < 1$ require care because all integrals must be regulated in a translationally invariant manner; $n=1$ is QCD; and theories with $n>1$ are straightforward because they are ultraviolet finite, \emph{i.e}., superrenormalisable.  Infrared divergences are absent because of dynamical chiral symmetry breaking.}
\begin{equation}
\label{Eqndependence}
{\mathpzc u}_n^\pi(x;\zeta) \stackrel{x\simeq 1}{\sim} (1-x)^{2n}.
\end{equation}

These remarks highlight that experimental validation of Eq.\,\eqref{pionDFpQCD} is a stern test of QCD and its role as the theory of strong interactions within the SM.  For that reason, we restate the equation in words:\\[0.4ex]
\hspace*{0.8\parindent}\parbox[t]{0.9\linewidth}{
{\sf T1}:
If QCD describes the pion, then at any scale for which an analysis of data using known techniques is valid, the form extracted for the pion's valence-quark DF must behave as $(1-x)^{\beta}$, $\beta>2$, on $x\gtrsim 0.9$ \cite{Holt:2010vj, Ball:2016spl, Courtoy:2020fex, Roberts:2021nhw}.
A result with $\beta<2$ entails one of the following: [\emph{a}] the analysis is incomplete, omitting or misrepresenting some aspect or aspects of the processes involved; [\emph{b}] (some of) the data being considered are not a true expression of a quality intrinsic to the pion; or [\emph{c}] QCD, as it is currently understood, is not the theory of strong interactions.}

%\medskip

%\noindent\emph{3.$\;$Experiments and the pion valence-quark distribution} ---
\section{Experiments and the pion valence-quark distribution}
\label{SecExpt}
The Drell-Yan process \cite{Drell:1970wh, Peng:2016ebs} $\pi + A \to \ell^+ \ell^- + X$, where $\ell$ is a lepton, $A$ is a nuclear target, and $X$ denotes the debris produced by the deeply inelastic reaction, has been used as the basis for all existing attempts to infer the large-$x$ behaviour of ${\mathpzc u}^\pi(x;\zeta)$ from experiment \cite{Corden:1980xf, Badier:1983mj, Betev:1985pg, Conway:1989fs}.
Analysing that reaction at leading order in pQCD, two terms appear in the cross-section \cite{Farrar:1975yb, Berger:1979du}.
One is directly sensitive to the pion LFWF and contributes the Eq.\,\eqref{pionDFpQCD} piece to the transverse part of the overall structure function.
The other is generated by what may be described as an initial state interaction between the valence antiquark in the pion and a valence quark in the target.  It is expressed as a $(1-x)^0\langle k_\perp^2\rangle / \zeta^2$ (higher-twist) longitudinal term in the overall cross-section, where $\langle k_\perp^2\rangle\ll\zeta^2$ measures the mean transverse momentum of the annihilating valence-antiquark and \mbox{-quark}.
Evidently, in any realisable experiment, there will always be a neighbourhood $x\simeq 1$ whereupon the longitudinal cross-section exceeds the transverse and access to ${\mathpzc u}^\pi(x;\zeta)$ is obscured.

%as $H_\pi^{\mathpzc u}(x,\xi,t;\zeta)$, then

%\cite{Ezawa:1974wm}, not all of which are rigorously linked to QCD, but the result is always the same; namely,

%one must distinguish between the pion valence-quark DF and the structure function measurement used to extract it, which contains another term that, for a given value of $\zeta$, can mask the DF piece on some domain of $x$ that \cite{Mueller:1981sg}.

%The DF is a property of the pion wave function whereas the structure function also expresses reaction specific effects.

%Structure function and Drell-Yan and each is associated with a different angular distribution in the lepton-pair rest frame.  This is not connected with the pion wave function but with the lepton pair production process.  Feature of scaling and non-scaling in the process, not in the meson wave function.

These remarks are important because existing attempts to extract the large-$x$ behaviour of the pion valence-quark DF \cite{Aicher:2010cb, Novikov:2020snp, Han:2020vjp, Barry:2021osv} are dominated by the data reported in Ref.\,\cite[E615]{Conway:1989fs}.  The angular distributions associated with that set and with the data in Ref.\,\cite[NA10]{Guanziroli:1987rp} show signs of the higher-twist contribution \cite{Wijesooriya:2005ir}.  However, extant analyses aimed at determining ${\mathpzc u}^\pi$ do not remove this contamination.  In addition, the binning of E615 data is not described in Ref.\,\cite{Conway:1989fs}; and this may introduce substantial uncertainty at large $x$.  For these reasons and also simply because experimental capabilities and techniques have improved in the thirty years since the E615 experiment, new data with much improved precision are essential if Eq.\,\eqref{pionDFpQCD} is to be properly tested.  Such measurements are in train \cite{JlabTDIS1, Adams:2018pwt, Aguilar:2019teb, Chen:2020ijn, Anderle:2021wcy, Arrington:2021biu}.

%\medskip

%\noindent\emph{4.$\;$Analyses of E615 data} ---
\section{Analyses of Drell-Yan data}
\label{SecFits}
\begin{figure}[t]
\includegraphics[width=0.435\textwidth]{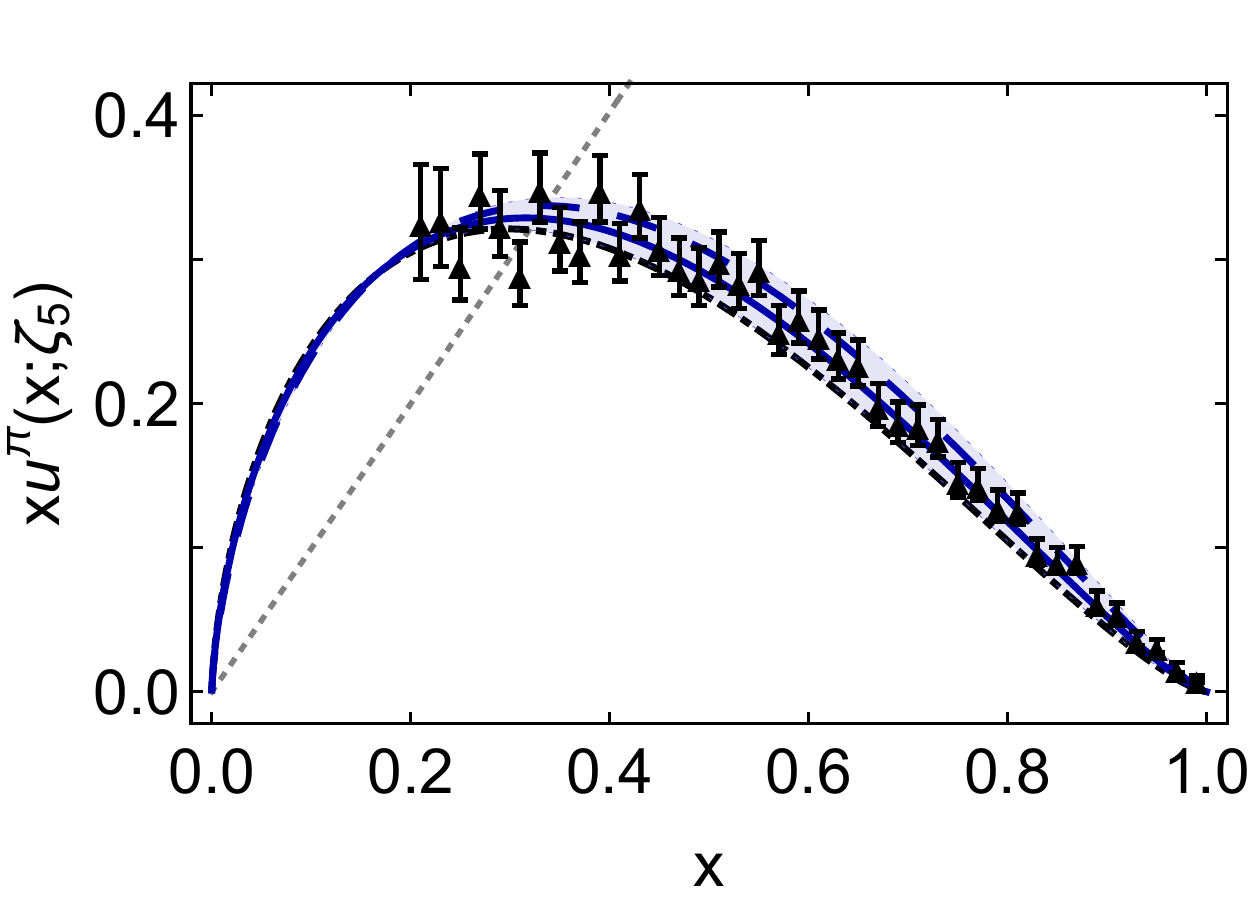}
\caption{\label{pionDF}
SCI Pion valence-quark DF, $x {\mathpzc u}_{\rm SCI}^\pi(x;\zeta_5=5.2\,{\rm GeV})$, obtained via evolution of the hadron scale DF in Eq.\,\eqref{SCIupizH}: $x {\mathpzc u}_{\rm SCI}^\pi(x;\zeta_{\cal H})$ is drawn as the dotted grey curve.
Evolution from:
$\zeta=\zeta_H=m_G$, Eq.\,\eqref{EqmG} -- dot-dashed black curve, using which $\chi^2/{\rm datum}=2.7$;
$\zeta=1.05 \zeta_H$ -- solid blue curve, $\chi^2/{\rm datum}=1.0$;
$\zeta=1.10 \zeta_H$ -- long-dashed blue curve, $\chi^2/{\rm datum}=1.0$;
and
$\zeta=1.15 \zeta_H$ -- upper edge of shaded blue band, $\chi^2/{\rm datum}=2.2$;
Data -- black up-triangles: described in Ref.\,\cite[E615]{Conway:1989fs} as measured values for the pion valence structure function.
}
\end{figure}

Data described as \cite[E615]{Conway:1989fs} ``Measured values for the pion valence structure function'', obtained via a leading-order (LO) pQCD analysis of Drell-Yan measurements and associated with a scale $\zeta=\zeta_5=5.2\,$GeV, are presented in Fig.\,\ref{pionDF}.  The set was used in Ref.\,\cite{Conway:1989fs} to argue that ${\mathpzc u}^\pi(x;\zeta_5)\sim (1-x)^{1.26(4)}$.  This is a marked contradiction of Eq.\,\eqref{pionDFpQCD}, a fact highlighted in Ref.\,\cite{Holt:2010vj}.

Working with the same data, a subsequent next-to-leading-order (NLO) pQCD analysis \cite{Aicher:2010cb}, which also included next-to-leading logarithm (NLL) threshold resummation effects (soft gluon resummation) \cite{Sterman:2000pt, Mukherjee:2006uu}, arrived at the result ${\mathpzc u}^\pi(x \simeq 1;\zeta_4=4\,{\rm GeV})\sim (1-x)^{2.34(7)}$, a markedly different outcome that is consistent with Eq.\,\eqref{pionDFpQCD}.

More recently, the effects of threshold resummation in analyses of the E615 data were reconsidered \cite{Barry:2021osv}, with comparison of three different methods for implementing such effects.
Two of the methods, which lie within the class of Mellin-Fourier schemes \cite{Sterman:2000pt, Mukherjee:2006uu} and may be called Mellin-Fourier-cosine (MFc) and Mellin-Fourier-expansion (MFe),  confirmed the outcome in\linebreak Ref.\,\cite{Aicher:2010cb}, returning a large-$x$ exponent $\beta(\zeta_c)>2$ at the scale $\zeta_c=1.27\,$GeV.
%% ... checked and this statement seems to be false = (In fact, in relation to contemporary predictions \cite{Ding:2019qlr, Ding:2019lwe, Cui:2021sfu, Cui:2020tdf}, the value of $\beta(\zeta_c)$ is surprisingly large.)
%
The third procedure, described as double-Mellin (dM) \cite{Westmark:2017uig}, produced an effective large-$x$ exponent \cite[Eq.\,(5)]{Barry:2021osv} $\beta_{\rm eff}(\zeta_c) \sim 1.2$, albeit with large uncertainty.
This disagreement admits the interpretation that E615 data and available methods for its analysis cannot are not able to deliver an unambiguous result for $\beta$; hence, cannot be used to test Eq.\,\eqref{pionDFpQCD}.
This, again, is a call for new, precise data.

%%You might perhaps like this a bit more: I have generated with a Monte Carlo 200 curves according to the chi^2-distribution defined with the E615 original data, at the experimental scale, and evolved them back to zetaH under the assumption of effective charge existence (no free parameter then).

%\medskip

%\noindent\emph{5.$\;$Content of E615 data} ---
\section{Content of Drell-Yan data}
\label{SecDYdata}
Given the situation described above, we judge it worth considering the E615 data from a different perspective; hence, ask the following question:\\[0.4ex]
\hspace*{0.8\parindent}\parbox[t]{0.90\linewidth}{
{\sf Q1}:
Suppose one has in hand a body of data which is truly an expression of pion structure and that a complete analysis of that data returns a valence-quark DF whose character is fairly represented by the points drawn in Fig.\,\ref{pionDF}, then what is the associated pion LFWF and what quark+antiquark interaction produced it?
}

\smallskip

\noindent Regarding the data in Fig.\,\ref{pionDF}, whose large-$x$ behaviour is known to be approximately linear \cite[E615]{Conway:1989fs}, the beginning of an answer is suggested by the discussion associated with Eq.\,\eqref{Eqndependence}: this DF can be connected with a vector$\,\times\,$vector quark+antiquark interaction that is momentum-independent, \emph{viz}.\ $[1/k^2]^{n\simeq 0}$.  In fact, this has been known since the first model calculations of Nambu-Goldstone boson valence DFs \cite{Davidson:1994uv}.

To elucidate, we consider the valence-quark DF for a $m_\pi=0.14\,$GeV pion obtained using a symmetry-preserving regularisation of a vector$\,\times\,$vector contact interaction (SCI) \cite{Zhang:2020ecj}:
\begin{equation}
\label{SCIupizH}
{\mathpzc u}_{\rm SCI}^\pi(x;\zeta_{\cal H}) = 1.007 - 0.0202 (1-2x)^2\,.
\end{equation}
%= 1.0145620024605506 - 0.020373898776417022*(1 - 2*x)^2
% = 1.0067389333434769 - 0.020216800030430453 (1-2x)^2
Since it provides a largely algebraic framework, the SCI has been widely used
\cite{Zhang:2020ecj, Yin:2021uom, Lu:2021sgg, Xu:2021mju, Xu:2021iwv} to set benchmarks for more complex studies using QCD-kindred interactions.  Comparisons with such calculations
\cite{Raya:2021zrz, Qin:2019hgk, Xu:2021mju, Yao:2021pdy} and also data \cite{Mokeev:2015lda, Segovia:2019jdk, Raya:2021pyr} show that SCI results are typically a reliable guide for long-wavelength properties of hadrons, such as masses and decay constants, but, unsurprisingly, fail in applications that are sensitive to and/or reveal structural properties.

The hadron-scale DF in Eq.\,\eqref{SCIupizH} is symmetric about $x=1/2$ because, by construction and implementation, dressed-quark degrees-of-freedom carry all properties of the bound-state at this scale: before gluon radiation begins, all glue and sea partons are sublimated into the dressed quarks.  As a consequence, one obtains
\begin{equation}
\label{EqSCI}
\langle x {\mathpzc u}_{\rm SCI}^\pi(x;\zeta_{\cal H}) \rangle
= \int_0^1 dx\, x\, {\mathpzc u}_{\rm SCI}^\pi(x;\zeta_{\cal H}) = \tfrac{1}{2}\,,
\end{equation}
\emph{viz}.\ each of the valence-quarks carries half the pion's light-front momentum at this scale.  (Isospin symmetry is assumed.)

The same qualities are expressed in every hadron-scale pseudoscalar-meson valence-quark distribution\linebreak computed using a framework that respects Poincar\'e covariance and QCD's vector and axial-vector Ward-Green-Takahashi identities, especially the pattern and consequences of dynamical chiral symmetry breaking (DCSB) \cite{Chang:2014lva, Ding:2019qlr, Ding:2019lwe}.  Any calculation that delivers an asymmetric DF at $\zeta_{\cal H}$ corresponds, implicitly or explicitly, to a treatment of the matrix element that defines the valence-quark distribution which violates at least one of these key QCD symmetry constraints.
(If the meson is built using nondegenerate valence degrees-of-freedom, $q_1$, $\bar q_2$, the only changes are $\langle x [{\mathpzc q}_1(x;\zeta_{\cal H})+\bar {\mathpzc q}_2(x;\zeta_{\cal H})]\rangle = 1$ and ${\mathpzc q}_1(x;\zeta_{\cal H})=\bar {\mathpzc q}_2(1-x;\zeta_{\cal H})$.)

In model calculations, following Ref.\,\cite{Jaffe:1980ti}, the value of $\zeta_{\cal H}$ is held to be a free parameter.  Its value is determined \emph{a posteriori} by requiring that, after using pQCD's evolution equations \cite{Dokshitzer:1977sg, Gribov:1971zn, Lipatov:1974qm, Altarelli:1977zs} to map the DF to another scale $\zeta=\zeta_E > \zeta_{\cal H}$, where ``empirical'' information about the DF is available, agreement is obtained with some chosen piece of that $\zeta=\zeta_E$ information.  There is an obvious yet typically overlooked objection to this procedure; namely,\\[0.4ex]
\hspace*{0.8\parindent}\parbox[t]{0.9\linewidth}{
%
%{\sf Caveat~A}.
{\sf O1}.
DGLAP evolution expresses properties intrinsic to four-dimensional QCD.  It is logically inconsistent to impose QCD-specific gluon radiation and splitting on a DF obtained from a quark+antiquark interaction that has no discernible link with QCD dynamics.}
\smallskip

\noindent If this caveat is ignored, then one merely arrives at a practitioner-preferred DF fit to the empirical $\zeta=\zeta_E$ information which supersedes its $\zeta=\zeta_H$ origin.

It is now worth observing that when considering model calculations which produce a pion-like pseudo\-sca\-lar-meson valence-quark DF, ${\mathpzc u}_{\mathpzc m}^\pi(x)$,  that is not symmetric around $x=1/2$, so that $\langle x {\mathpzc u}_{\mathpzc m}^\pi(x) \rangle < 1/2$, it is logically consistent to interpret the result as corresponding to a model scale $\zeta = \zeta_{\mathpzc m} > \zeta_{\cal H}$, so long as the model ensures that physical observables are independent of this choice of scale.  (If the latter is not true, then the model can be improved so this weakness is remedied.)  Then, since the model's proponent is usually prepared to use pQCD DGLAP evolution to map the DF to a higher scale, it is equally acceptable to use the same procedure and work backwards to determine the scale, $\zeta_{\cal H}$, at which the model's valence-quarks saturate the momentum sum rule: $\langle 2 x {\mathpzc u}_{\mathpzc m}^\pi(x;\zeta_{\cal H}) \rangle =1$.
(So evolved, the DF might also be symmetric around $x=1/2$, in which case the procedure delivers a $\zeta=\zeta_{\cal H}$ DF that is implicitly corrected for the symmetry violation(s) in the model's formulation.)
It is now plain that all model studies support the same definition of $\zeta_{\cal H}$.

In the QCD context, distinct from both modelling and data fitting, it is possible to predict a unique value of $\zeta_H$ based on the behaviour of QCD's process-independent (PI) effective charge \cite{Binosi:2016nme, Cui:2019dwv}, $\hat{\alpha}(k^2)$.  This charge agrees to better than 0.1\% with pQCD's one-loop coupling on $k^2 \gtrsim 4 m_p^2$.  However, as $k^2$ continues to run toward zero, $\hat{\alpha}(k^2)$ exhibits a qualitative change; and its behaviour on
\begin{equation}
\label{EqmG}
k^2 < m_G^2 := (0.331(2)\,{\rm GeV})^2\,,
\end{equation}
indicates that gauge sector modes with $k^2 \lesssim m_G^2$ are screened from interactions and QCD has entered a practically conformal domain.
The value of $m_G$ is determined by $m_0$, QCD's renormalisation group invariant gluon mass scale \cite{Binosi:2016nme, Cui:2019dwv, Roberts:2021nhw}.
It follows that the line $k^2=m_G^2$ separates long- and short-wavelength physics; hence, serves as the natural definition for the hadron scale, \emph{viz}.\ $\zeta_H=m_G$, because, below such momenta, gauge sector modes have effectively decoupled and gluon emission is frozen out.  This is the position introduced in Refs.\,\cite{Ding:2019qlr, Ding:2019lwe, Cui:2019dwv, Cui:2020dlm, Cui:2020tdf, Chang:2021utv}, which further argue that evolution should be implemented by using $\hat{\alpha}(k^2)$ to integrate the one-loop DGLAP equations.  Additional features of this all-orders evolution scheme are described elsewhere \cite[Sec.\,VII]{Raya:2021zrz}.   Naturally, given the properties of $\hat{\alpha}(k^2)$, the approach is equivalent to standard DGLAP evolution on any domain upon which pQCD is applicable.

Neglecting {\sf O1} for the moment, then the all-orders evolution scheme yields the dot-dashed black curve in Fig.\,\ref{pionDF}.  This prediction is made without reference to E615 data so the $\chi^2/{\rm datum}=2.7$ outcome is striking.
%% chi2 = 2.73
%% zH(1.05) ch2=1.04
As shown in Fig.\,\ref{pionDF}, supposing only that the prediction for $\zeta_{\cal H}$ in Eq.\,\eqref{EqmG} has a 5\% uncertainty, leads to a description of the data with $\chi^2/{\rm datum}=1.0$.

Returning to the question posed at the beginning of this section, {\sf Q1}, the outcomes just described invite the following answer: A valence-quark DF whose character is fairly represented by the points drawn in Fig.\,\ref{pionDF} is derived from a pion LFWF that is produced by a momentum-independent quark+antiquark interaction.
However, this position is untenable because DGLAP evolution cannot be derived from such an interaction.
%then by the principle of \emph{reductio ad absurdum},
%However, that answer is absurd because DGLAP evolution cannot be derived from such an interaction.
Logically, therefore, the data in Fig.\,\ref{pionDF} cannot be a true representation of the valence-quark DF of a pion generated by QCD interactions. Thus, either the analysis leading to such data was incomplete or the experimental data on which the analysis was based is not a veracious expression of an intrinsic property of the pion, or both issues have played a role.

%\medskip

%\noindent\emph{6.$\;$New look at E615 data} ---
%\noindent\emph{6.$\;$Character and consequences of all orders evolution} ---
\section{Character and consequences of all orders evolution}
\label{SecAOE}
\begin{figure}[t]
\includegraphics[width=0.435\textwidth]{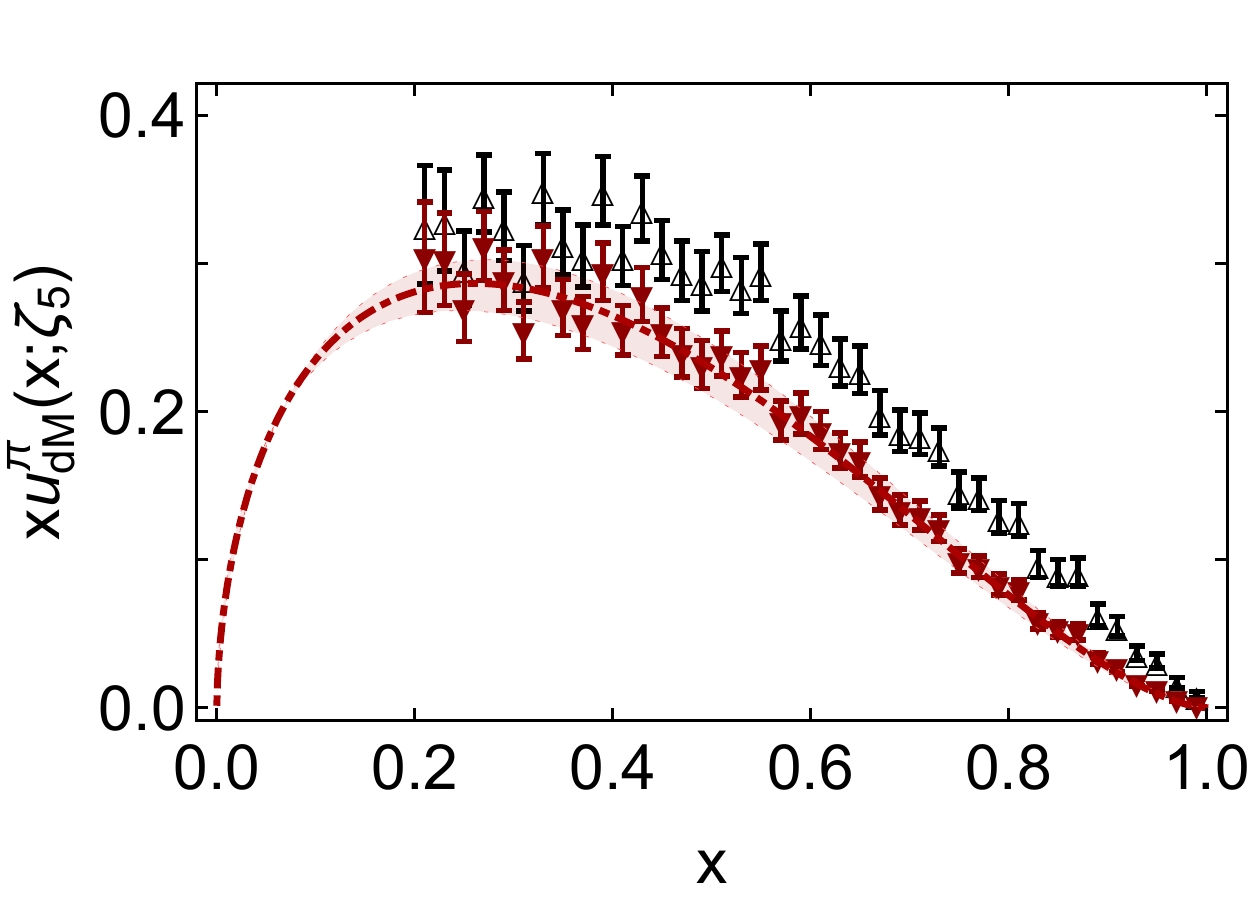}
\caption{\label{FdME615}
Pion valence-quark DF (dot-dashed red curve) inferred from the fit to data in Ref.\,\cite{Barry:2021osv} that was obtained with NLL threshold resummation performed using the dM method, evolved $\zeta_c \to \zeta_5$.
[The encompassing band expresses the uncertainty in Eqs.\,\eqref{dMupizc}.]
At this scale, the large-$x$ exponent is $\beta \approx 1.5$.
Data recorded in Ref.\,\cite[E615]{Conway:1989fs} -- black open up-triangles;
and red down-triangles -- projection of that data onto the $\zeta_c \to \zeta_5$ evolution of Eq.\,\eqref{dMupizc}, which we denote as E615$_{\rm dM}$.
}
\end{figure}

A dM approach to including NLL threshold resummation in the analysis of Drell-Yan data is described in Ref.\,\cite{Barry:2021osv}.  Based on the information therein, we have developed the following approximate form for the extracted valence-quark DF:
\begin{subequations}
\label{dMupizc}
\begin{align}
 &{\mathpzc u}_{\rm dM}^\pi(x;\zeta_c)  = n_{\mathpzc u} x^\alpha (1-x)^{\beta} (1+\gamma x^2)\,,\\
 &\alpha = -0.40_{\mp 0.05}\,,\;
 \beta = 1.23_{\pm 0.05} \,,\;
 \gamma = 0.62_{\pm 0.22}  \,,  \label{E615JAMparams}
\end{align}
\end{subequations}
where $n_{\mathpzc u}$ ensures unit normalisation.  Fig.\,\ref{FdME615} depicts the E615 data from Fig.\,\ref{pionDF} projected onto the central curve in Eq.\,\eqref{dMupizc} after its evolution $\zeta_c \to \zeta_5$.  The original and new analyses are qualitatively equivalent.

On the other hand, however, there is a key quantitative difference between the original and new analysis, \emph{viz}.\ the Ref.\,\cite{Barry:2021osv} fits lodge less of the pion's light-front momentum with the valence quarks: at $\zeta_c$, the fraction is $0.46(3)$.  This value is significantly smaller than the prediction $0.52(2)$ that is typical of contemporary continuum and lattice QCD calculations at this scale \cite{Cui:2020tdf, Chang:2021utv}.

Regarding the E615$_{\rm dM}$ data in Fig.\,\ref{FdME615}, it is plain that a repetition of the analysis in Sec.\,5 would yield a qualitatively equivalent outcome to that exposed by our discussion of the original E615 data.  We will therefore follow a more general approach based on a single proposition:\\[0.4ex]
\hspace*{0.8\parindent}\parbox[t]{0.9\linewidth}{
{\sf P1}: There exists an effective charge, $\alpha_{1\ell}(k^2)$, that, when used to integrate the one-loop pQCD DGLAP equations, defines an evolution scheme for parton DFs that is all-orders exact.  $\alpha_{1\ell}(k^2)$ need not be unique.
}

\smallskip

\noindent It is here worth recording some background and corollaries of this proposition.

A context for {\sf P1} is provided by the discussion of process-dependent charges in Refs.\,\cite{Grunberg:1982fw, Grunberg:1989xf, Dokshitzer:1998nz}; so, this charge need not be process independent, \emph{e.g}., different charges may be needed for distinct observables.  On the other hand, a process-independent charge with this character is not excluded, as evidenced by the efficacy of that discussed in Refs.\,\cite{Binosi:2016nme, Cui:2019dwv, Cui:2020dlm, Cui:2020tdf, Chang:2021utv}.

As detailed elsewhere \cite{Deur:2016tte}, in being defined by an observable (here, a structure function) $\alpha_{1\ell}(k^2)$ is consistent with the renormalisation group.  It is also renormalisation scheme independent; everywhere finite and analytic; and supplies an infrared completion of any standard running coupling.

As a consequence of finiteness, {\sf P1} entails that there is a scale, which we have denoted $\zeta_{\cal H}$ above, such that
\begin{equation}
\label{twoxone}
\langle 2 x {\mathpzc u}^\pi(x;\zeta_{\cal H})\rangle = 1\,,
\end{equation}
\emph{viz}.\ whereat the $\zeta=\zeta_{\cal H}$ dressed valence quarks carry all the pion's light-front momentum.  Momentum conservation now demands that the glue and sea light-front momentum fractions vanish at $\zeta_{\cal H}$; and since all physical DFs are nonnegative on $x\in[0,1]$, it follows that the glue and sea DFs vanish identically at this scale.

Eq.\,\eqref{twoxone} is guaranteed for any symmetric function, \emph{i.e}., when
\begin{equation}
\label{EqSymm}
{\mathpzc u}^\pi(x;\zeta_{\cal H}) = {\mathpzc u}^\pi(1-x;\zeta_{\cal H})\,.
\end{equation}
This is a sufficient but not necessary condition.
However, as noted after Eq.\,\eqref{EqSCI}, Eq.\,\eqref{EqSymm} is satisfied by the result of any calculation which respects Poincar\'e covariance and QCD's vector and axial-vector Ward-Green-Takahashi identities, particularly the pattern and consequences of DCSB \cite{Chang:2014lva, Ding:2019qlr, Ding:2019lwe}.  A violation of Eq.\,\eqref{EqSymm} by a valence-quark DF obtained by whatever means links this DF to a treatment of the underlying matrix element that breaks basic QCD symmetries.

As noted above, when considering a pseudoscalar meson built using nondegenerate valence degrees-of-freedom, $q_1$, $\bar q_2$, the only changes are $\langle x [{\mathpzc q}_1(x;\zeta_{\cal H})+\bar {\mathpzc q}_2(x;\zeta_{\cal H})]\rangle = 1$ and ${\mathpzc q}_1(x;\zeta_{\cal H})=\bar {\mathpzc q}_2(1-x;\zeta_{\cal H})$.

Now, explicitly describing the isospin symmetry limit, consider the Mellin moments:
\begin{equation}
\label{EqMellin}
\langle x^n \rangle_{{\mathpzc u}_\pi}^{\zeta_{\cal H}}
:= \langle x^n  {\mathpzc u}^\pi(x;\zeta_{\cal H})\rangle
= \int_0^1 dx\,x^n\, {\mathpzc u}^\pi(x;\zeta_{\cal H})\,.
\end{equation}
As highlighted in Ref.\,\cite[Eq.\,(29)]{Ding:2019lwe}, Eq.\,\eqref{EqSymm} entails that for $n\in {\mathbb Z}^{\geq}$, $\langle x^{2n+1} \rangle_{{\mathpzc u}_\pi}^{\zeta_{\cal H}}$ is linearly dependent on (completely determined by) the set of even moments $\langle x^{2m} \rangle_{{\mathpzc u}_\pi}^{\zeta_{\cal H}}$ with $m\leq n$.  For instance,
{\allowdisplaybreaks
\begin{subequations}
\label{EqExplictRecursion}
\begin{align}
\langle x^1 \rangle_{{\mathpzc u}_\pi}^{\zeta_{\cal H}} & = \tfrac{1}{2} \langle x^0 \rangle_{{\mathpzc u}_\pi}^{\zeta_{\cal H}}\,, \label{line1} \\
\langle x^3 \rangle_{{\mathpzc u}_\pi}^{\zeta_{\cal H}} & =
\tfrac{1}{2} \langle x^0 \rangle_{{\mathpzc u}_\pi}^{\zeta_{\cal H}}
- \tfrac{3}{2} \langle x^1 \rangle_{{\mathpzc u}_\pi}^{\zeta_{\cal H}}
+ \tfrac{3}{2} \langle x^2 \rangle_{{\mathpzc u}_\pi}^{\zeta_{\cal H}} \nonumber \\
& \stackrel{\mbox{\footnotesize Eq.\,\eqref{line1}}}{=}
-\tfrac{1}{4} \langle x^0 \rangle_{{\mathpzc u}_\pi}^{\zeta_{\cal H}}
+ \tfrac{3}{2} \langle x^2 \rangle_{{\mathpzc u}_\pi}^{\zeta_{\cal H}}\,,
% \\
%
%{\rm etc}. & \nonumber
\end{align}
\end{subequations}
etc.  These constraints can most straightforwardly be exploited using the following recursion relation for the odd-order Mellin moments:
\begin{align}
\langle & x^{2 n+1}\rangle_{{\mathpzc u}_\pi}^{\zeta_{\cal H}}\nonumber  \\
& = \frac{1}{2(n+1)}\sum_{j=0,1,\ldots}^{2n}(-)^j
\left(\begin{array}{c} 2(n+1) \\ j \end{array}\right) \langle x^j\rangle_{{\mathpzc u}_\pi}^{\zeta_{\cal H}}\,.
\label{EqSymmB}
\end{align}
}

Furthermore, given {\sf P1}, then \cite[Sec.\,VII]{Raya:2021zrz}:
\begin{equation}
\label{EqxnzzH}
\langle x^n\rangle_{{\mathpzc u}_\pi}^\zeta
= \langle x^n\rangle_{{\mathpzc u}_\pi}^{\zeta_{\cal H}}
\left( \langle 2 x \rangle_{{\mathpzc u}_\pi}^\zeta\right)^{\gamma_0^n/\gamma_0^1},
\end{equation}
where, for $n_f=4$ quark flavours,
\begin{equation}
\label{eq:Dfn}
\gamma_0^n = -\frac 4 3 \left( 3 + \frac{2}{(n+1)(n+2)} - 4 \sum_{j=1}^{n+1} \frac 1 j  \right) \,.
\end{equation}
Namely, given the pion valence-quark DF at one scale, here identified as $\zeta_{\cal H}$, then its full $x$-dependence at any other scale $\zeta$ is completely determined by the value of its first Mellin moment at $\zeta$.  No other knowledge is needed, including and especially not information on the form of $\alpha_{1\ell}(k^2)$.
Inserting Eq.\,\eqref{EqxnzzH} into Eq.\,\eqref{EqSymmB}, one finds
\begin{align}
\langle & x^{2 n+1}\rangle_{{\mathpzc u}_\pi}^{\zeta}
= \frac{(\langle 2 x \rangle_{{\mathpzc u}_\pi}^\zeta)^{\gamma_0^{2n+1}/\gamma_0^1}}{2(n+1)} \nonumber \\
& \times \sum_{j=0,1,\ldots}^{2n}(-)^j
\left(\begin{array}{c} 2(n+1) \\ j \end{array}\right)
\langle x^j\rangle_{{\mathpzc u}_\pi}^{\zeta}
(\langle 2 x \rangle_{{\mathpzc u}_\pi}^\zeta)^{-\gamma_0^{j}/\gamma_0^1}\,.
\label{RecursionAnyz}
\end{align}
Conversely, any DF whose moments fulfill this recursion relation is related by evolution to a symmetric function at $\zeta_{\cal H}$.

\begin{figure}[t]
\includegraphics[width=0.435\textwidth]{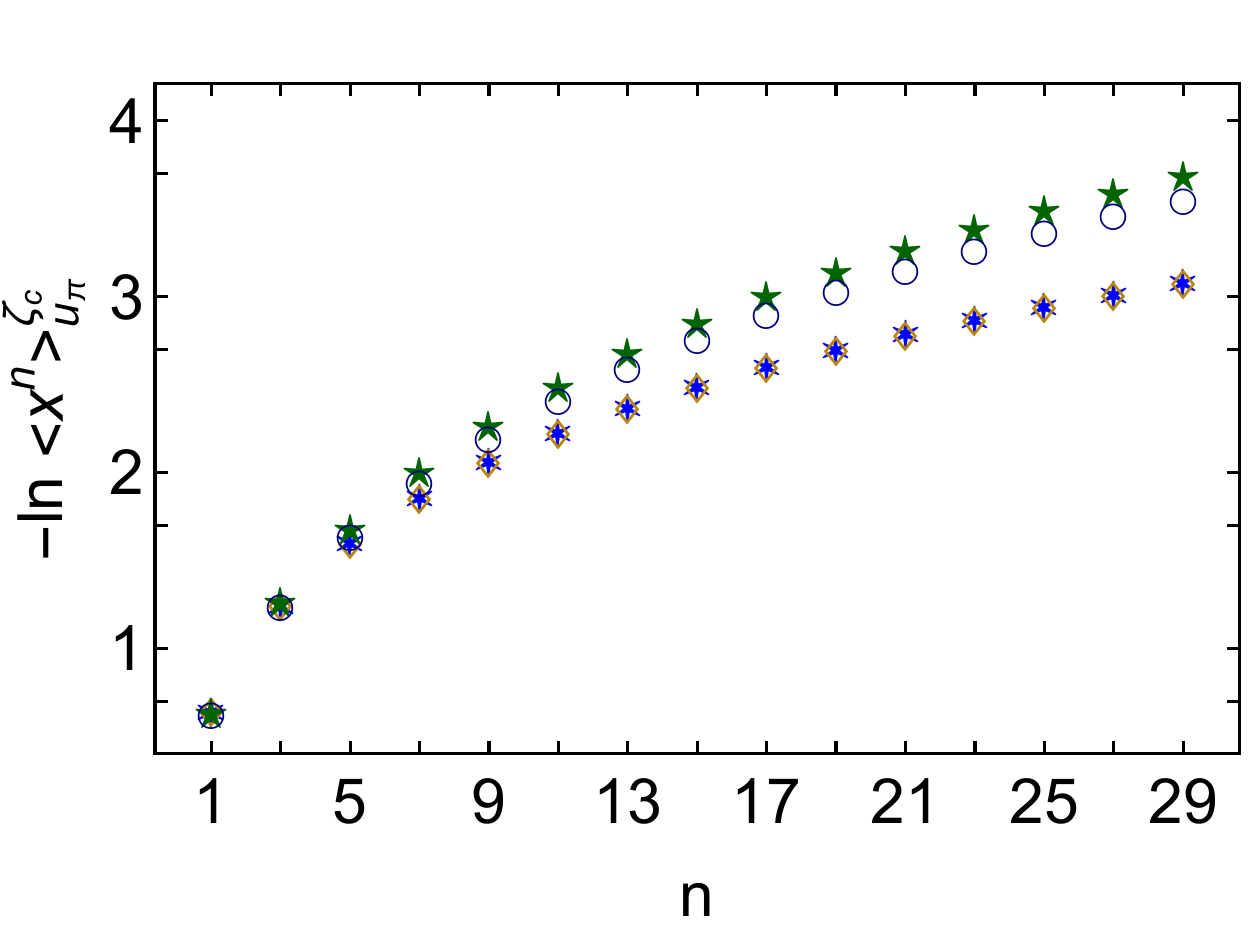}
\caption{\label{FigOddMoments}
Odd moments computed from the DF in Eq.\,\eqref{dMupizc}: directly, using Eq.\,\eqref{EqMellin} (blue asterisks); and using the recursion relation in Eq.\,\eqref{RecursionAnyz} (gold open diamonds).
Also, odd moments computed from the DF in Eq.\,\eqref{MFupizc}: directly, using Eq.\,\eqref{EqMellin} (green stars); and using the recursion relation in Eq.\,\eqref{RecursionAnyz} (navy-blue open circles).
Recall: the recursion relation is satisfied if, and only if, the source function is related by evolution to a DF that is symmetric at some scale $\zeta_{\cal H}$; evidently, this applies to Eqs.\,\eqref{dMupizc} but not Eqs.\,\eqref{MFupizc}.
}
\end{figure}

This corollary can be examined in connection with any inferred or calculated pion DF.  Hence, it can be tested with Eq.\,\eqref{dMupizc}.  Fig.\,\ref{FigOddMoments} displays the first $15$ odd moments computed directly from the central curve in Eq.\,\eqref{dMupizc} compared with the values for these moments predicted by Eq.\,\eqref{RecursionAnyz}: there is precise agreement.  (Given the factorial growth of the coefficients and exponential decrease in the moments, one must be careful to eliminate round-off error when evaluating the sum in Eq.\,\eqref{RecursionAnyz}.)

The agreement verified by Fig.\,\ref{FigOddMoments} entails that the DF in Eq.\,\eqref{dMupizc}, which fairly represents that extracted from a particular analysis of data relevant to the pion valence-quark DF \cite{Barry:2021osv}, is related by evolution to a symmetric DF at a scale $\zeta_{\cal H}<\zeta_c$.
As noted above, momentum conservation now demands that the glue and sea DFs must both vanish identically at this scale.
Consequently, when analysing data, one should not independently fit valence-quark, glue, and sea distributions at any scale $\zeta>\zeta_{\cal H}$ because the glue and sea distributions at $\zeta > \zeta_{\cal H}$ are completely determined by ${\mathpzc u}^\pi(x;\zeta_{\cal H})$ and the evolution equations: the glue and sea distributions are not independent functions.

It is also instructive to test Eq.\,\eqref{RecursionAnyz} using a calculated pion valence-quark DF instead of a fit to data.  In this connection, consider that Ref.\,\cite{Alexandrou:2021mmi} used lattice-QCD output to build a $x$-dependent pion valence-quark DF at $\zeta=\zeta_5$.  (Its large-$x$ behaviour is consistent with Eq.\,\eqref{pionDFpQCD} \cite[Fig.\,12]{Alexandrou:2021mmi}.)  The first six nontrivial Mellin moments of this DF are listed in the first column here, with the indicated statistical and systematic errors:
\begin{equation}
\begin{array}{l|cc}
& \multicolumn{2}{c}{\mbox{$\langle x^n \rangle_{{\mathpzc u}_\pi}^{\zeta_5}$}} \\
n &   \mbox{Ref.\,\cite{Alexandrou:2021mmi}} & \mbox{Eq.\,\eqref{RecursionAnyz}} \\\hline
1 &  0.230(3)(7)  & \underline{0.230}\phantom{8} \\
2 &  0.087(5)(8)  & \underline{0.087}\phantom{8} \\
3 &  0.041(5)(9)  & 0.041\phantom{8} \\
4 &  0.023(5)(6)  & \underline{0.023}\phantom{8}\\
5 &  0.014(4)(5)  & 0.015\phantom{8}\\
6 &  0.009(3)(3)  & \underline{0.009}\phantom{8}\\
7 &  & 0.0078
\end{array}\,.
\end{equation}
% 0.230(3)(7) 0.087(5)(8) 0.041(5)(9) 0.023(5)(6) 0.014(4)(5) 0.009(3)(3)
The second column reports the results computed from the first column using Eq.\,\eqref{RecursionAnyz}.  The underlined entries are the linearly independent moments, used as input, and the remaining entries are the Eq.\,\eqref{RecursionAnyz} predictions.
Notably, the linearly dependent $n=7$ moment is not reported in Ref.\,\cite{Alexandrou:2021mmi}; so, our result is a prediction.  Working with Ref.\,\cite[Eq.\,(46) and Table\,X]{Alexandrou:2021mmi}, one obtains $\langle x^7 \rangle_{{\mathpzc u}_\pi}^{\zeta_5}=0.0065(24)$.
The agreement between lattice calculation and the even-function recursion relation is striking.  It suggests that this lattice result, too, is linked via evolution to a symmetric DF at a scale $\zeta_{\cal H}<\zeta_5$.

These two examples illustrate the following general conclusion:\\[0.4ex]
\hspace*{0.8\parindent}\parbox[t]{0.9\linewidth}{
{\sf C1}: Given a pion valence-quark DF, ${\mathpzc u}^\pi(x;\zeta_{\cal E})$, either fitted to data or calculated at some scale $\zeta_{\cal E}$, whose moments satisfy Eq.\,\eqref{RecursionAnyz}, then there is always a $\zeta_{\cal H}\in[0,\zeta_{\cal E})$ such that evolution $\zeta_{\cal E} \to \zeta_{\cal H}$ maps ${\mathpzc u}^\pi(x;\zeta_{\cal E})\to {\mathpzc u}^\pi(x;\zeta_{\cal H})$, with ${\mathpzc u}^\pi(1-x;\zeta_{\cal H})={\mathpzc u}^\pi(x;\zeta_{\cal H})$.
At $\zeta_{\cal H}$, valence-quarks carry all the pion's light-front momentum and the glue and sea distributions vanish.
}

%\smallskip
%
%Given a valence-quark DF, ${\mathpzc u}_\pi(x;\zeta_{\cal C})$, fitted or calculated at some scale $\zeta_{\cal C}$, then there is always a $\zeta_{\cal H}<\zeta_{\cal C}$ such that evolution $\zeta_{\cal C} \to \zeta_{\cal H}$ maps ${\mathpzc u}^\pi(x;\zeta_{\cal C})\to {\mathpzc u}^\pi(x;\zeta_{\cal H})$, with ${\mathpzc u}^\pi(x;\zeta_{\cal H})={\mathpzc u}^\pi(1-x;\zeta_{\cal H})$.

%We find that the reconstruction is feasible and that our lattice data favor a large x-dependence that falls as (1-x)2 for both the pion and kaon PDFs. We integrate the reconstructed PDFs to extract the higher moments with 4≤n≤6. Finally, we compare the pion and kaon PDFs, as well as the ratios of their moments, to address the effect of SU(3) flavor symmetry breaking.

%\medskip

%\noindent\emph{6.$\;$New look at E615 data} ---
%\noindent\emph{7.$\;$Reviewing E615$_{\rm dM}$ data} ---
\section{Reviewing E615$_{\rm\bf dM}$ data}
\label{NLLdM}
Armed with {\sf P1} and its manifold consequences, we now consider what can additionally be revealed about the content of the E615$_{\rm dM}$ data in Fig.\,\ref{FdME615}.  (Sec.\,6 has already shown that the data fit is linked via evolution to a symmetric DF at some scale $\zeta_{\cal H}<\zeta_c$.)  We do not constrain ourselves to solely this set, however.  Instead, we consider an array of possibilities with the same qualitative character, \emph{viz}.\ a set representative of the outcome of any analysis of a body of data that returns a pion valence-quark DF for which the effective large-$x$ exponent is $\beta(\zeta_c) \approx 1.2$.

\begin{figure}[t]
\vspace*{2ex}

\leftline{\hspace*{0.5em}{\large{\textsf{A}}}}
\vspace*{-3ex}
\includegraphics[width=0.435\textwidth]{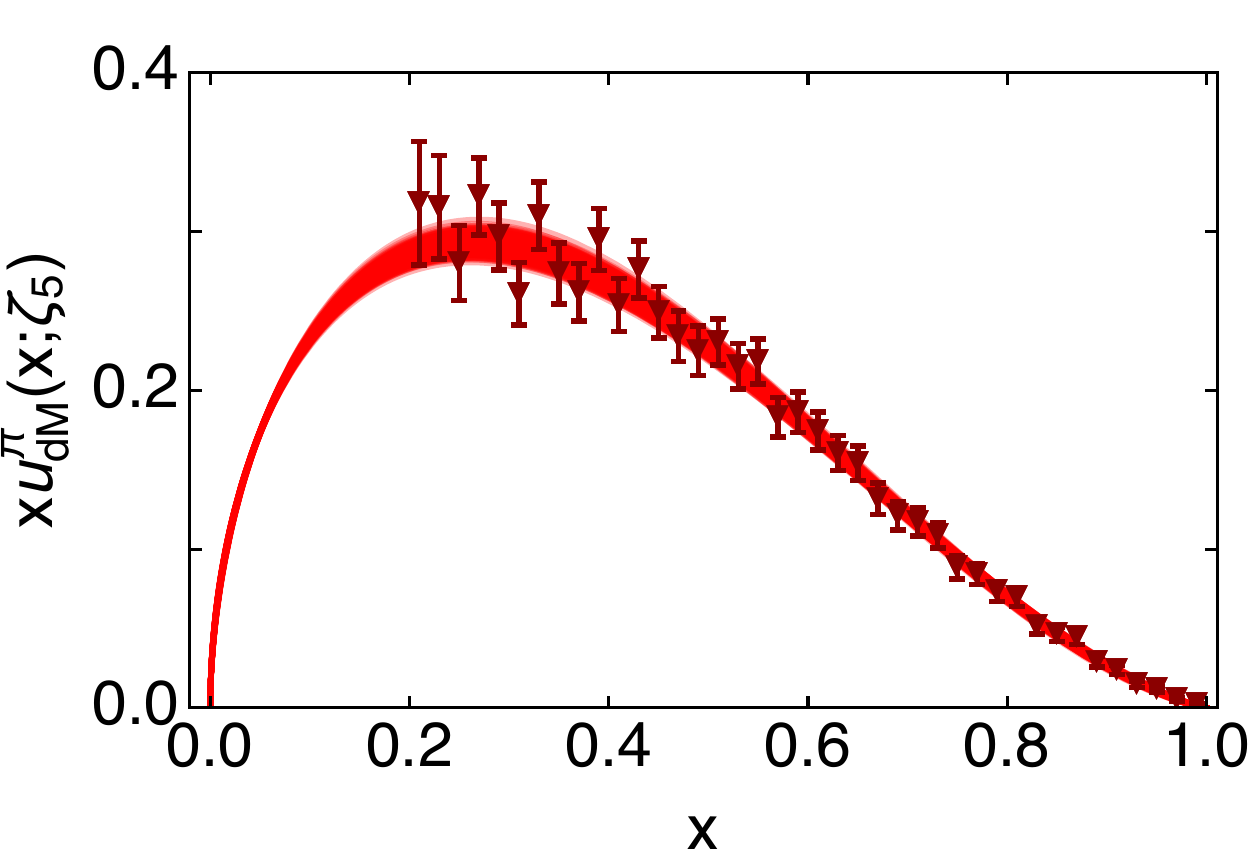}
\vspace*{-1ex}

\leftline{\hspace*{0.5em}{\large{\textsf{B}}}}
\vspace*{-3ex}
\includegraphics[width=0.435\textwidth]{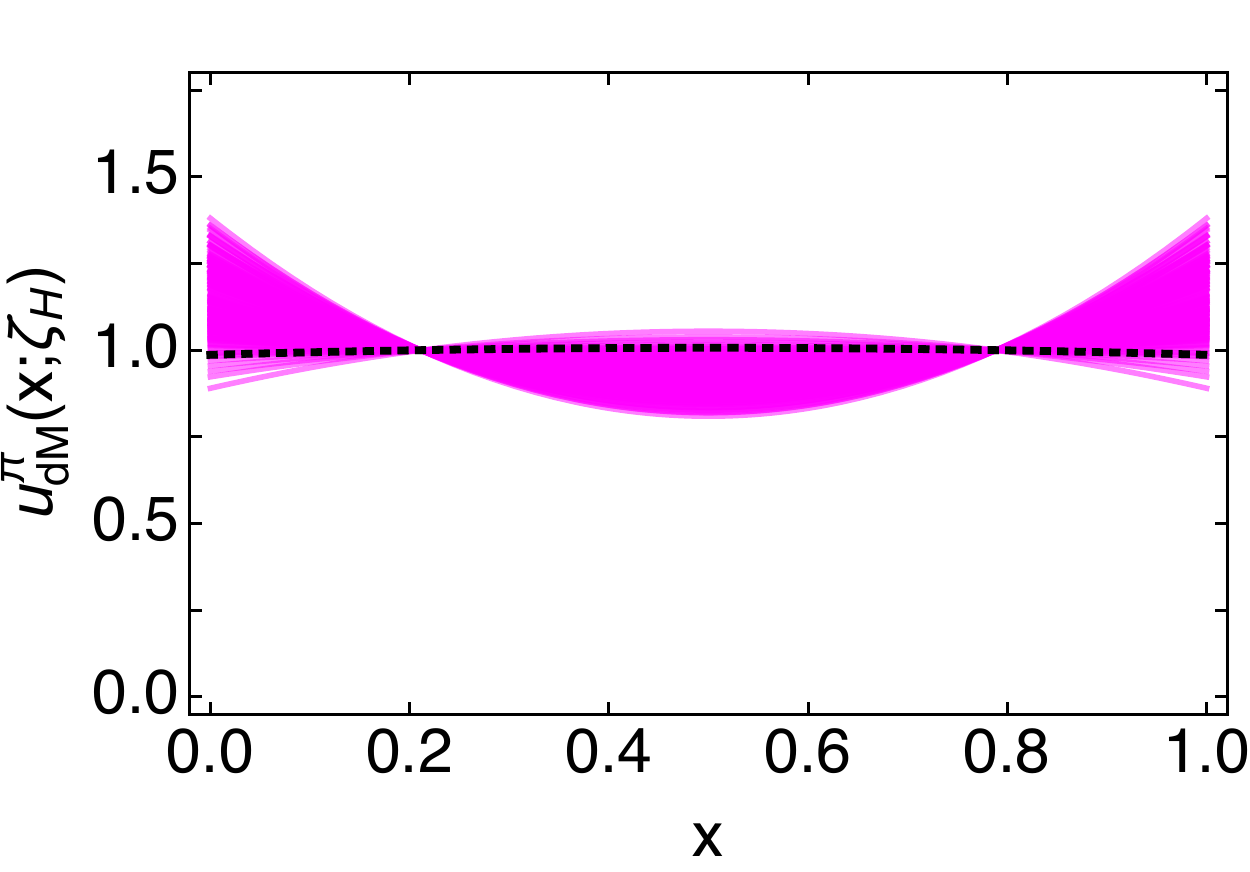}
\caption{\label{FdME615K}
\emph{Upper panel}\,--\,{\sf A}.  Randomly distributed ensemble of pion valence-quark DFs (red curves) constructed from the E615$_{\rm dM}$ data (red down-triangles) using the procedure described in connection with Eq.\,\eqref{EqProb}.
\emph{Lower panel}\,--\,{\sf B}.  $\zeta_5\to \zeta_{\cal H}$ evolution of each curve in Panel {\sf A} (magenta curves).  The SCI result in Eq.\,\eqref{SCIupizH} is drawn as the dotted black curve.
}
\end{figure}

To achieve this generalisation, we first suppose that a fair approximation to any such DF is provided by
\begin{equation}
\label{BestDF}
%{\mathpzc u}_\pi(x;[\alpha_i];\zeta) = {\mathpzc n}_\pi^\zeta
%x^{\alpha_1^\zeta} (1-x)^{\alpha_2^\zeta} (1 + \alpha_3^\zeta \surd x + \alpha_4^\zeta x)\,,
{\mathpzc u}^\pi(x;[\alpha_i];\zeta) = {\mathpzc n}_{\mathpzc u}^\zeta
x^{\alpha_1^\zeta} (1-x)^{\alpha_2^\zeta} (1 + \alpha_3^\zeta x^2)\,,
\end{equation}
where ${\mathpzc n}_{\mathpzc u}^\zeta$ ensures unit normalisation.  This is a weak assumption: any of the forms commonly used in fitting data would serve equally well.  Then, we proceed as follows.
(\emph{i}) Determine central values of $\{\alpha_i^\zeta|i=1,2,3\}$ via a least-squares fit to the $\zeta=\zeta_5$ E615$_{\rm dM}$ data.  %Verifying internal consistency, this reproduces the values in Eq.\,\eqref{E615JAMparams}.
(\emph{ii}) Generate a new vector $\{\alpha_i^\zeta|i=1,2,3\}$, each element of which is distributed randomly around its best-fit value.
(\emph{iii})
Using the DF obtained therewith, evaluate
\begin{equation}
\chi^2 = \sum_{l=1}^{N}\frac{({\mathpzc u}^\pi(x_l;[\alpha_i];\zeta_5)-u_j)^2}{\delta_l^2}\,,
\end{equation}
where $\{(x_l,u_l\pm\delta_l) | N=1,\ldots\,40\}$ is the E615$_{\rm dM}$ data set.  This $\{\alpha_i\}$ configuration is accepted with probability
\begin{equation}
\label{EqProb}
{\mathpzc P}  = \frac{P(\chi^2;d)}{P(\chi_0^2;d)} \,, \;
P(y;d) = \frac{(1/2)^{d/2}}{\Gamma(d/2)} y^{d/2-1} {\rm e}^{-y/2}\,,
\end{equation}
%%\begin{subequations}
%%\label{EqProb}
%%\begin{align}
%%{\mathpzc P} & = \frac{P(\chi^2;d)}{P(\chi_0^2;d)} \,, %  \leq 1\,,\\
%
%%P(y;d) & = \frac{(1/2)^{d/2}}{\Gamma(d/2)} y^{d/2-1} {\rm e}^{-y/2}\,,
%%\end{align}
%%\end{subequations}
where $d=N-3$ and $\chi_0^2 \approx d$ locates the maximum of the $\chi^2$-probability density, $P(\chi^2;d)$.
(\emph{iv}) Repeat (\emph{ii}) and (\emph{iii}) until one has a $K \gtrsim 1000$-member ensemble of DFs.
This procedure yields the collection of DFs drawn in Fig.\,\ref{FdME615K}A.

With any collection of pion DFs in hand, one also has associated ensembles of Mellin moments:\linebreak
$\{ {\cal M}_j(n;\zeta_5) | j = 1, \ldots, K\}$,
\begin{equation}
{\cal M}_j(n,\zeta_5) = \int_0^1 dx\, x^n\, {\mathpzc u}_\pi(x;[\alpha_i]^j;\zeta_5)\,,
\end{equation}
where $n \in {\mathbb Z}^{\geq}$ and $\{ [\alpha_i]^j | j=1,\ldots,K\}$ are the coefficient vectors associated with the accepted DFs.
Eq.\,\eqref{EqxnzzH} can be used to evolve each of these moments to any other scale, $\zeta$; and given that a continuous function of compact support is uniquely defined by its Mellin moments, one can therefrom reconstruct each of the distributions at $\zeta$.  Plainly, the mean and standard deviation of the $n^{\rm th}$ moment can be computed at the new scale.

Using Eq.\,\eqref{EqxnzzH}, we evolved each one of the $K$ replicas of the E615$_{\rm dM}$ fit in Fig.\,\ref{FdME615K} from $\zeta_5\to \zeta_{\cal H}$ with the results drawn in Fig.\,\ref{FdME615K}B: all the evolved curves are consistent with {\sf C1}.  Furthermore, comparison with the SCI result in Eq.\,\eqref{SCIupizH} shows that every one of the evolved curves is contained within the class of DFs linked to a momentum-independent quark+antiquark interaction.\footnote{The appearance of concave-up curves in the evolved set of replicas likely results from imprecision in the original binning of the E615 data, which introduces substantial additional uncertainty at large $x$ that is not expressed in the quoted errors.}

These outcomes add weight to the discussion in Sec.\,\ref{SecDYdata}.
Namely, the replicas in Fig.\,\ref{FdME615K} serve as a representative body of extracted forms for the pion's valence-quark DF whose large-$x$ behaviour is determined by the E615 Drell-Yan data.
Accepting {\sf P1}, then that data cannot be connected with a pion LFWF which is derived from a momentum-dependent quark+antiquark interaction; hence, the data cannot be a fair representation of the valence-quark DF of a pion generated by QCD interactions.
Consequently, either the analysis of data that formed the basis for the replicas was incomplete or the source Drell-Yan data itself does not truly express an intrinsic property of the pion, or both problems have interfered together.

%\smallskip

%\noindent\emph{8.$\;$Threshold resummation using cosine and expansion methods} ---
%
\section{Threshold resummation using Mellin-Fourier methods}
\label{NLLMF}
Threshold resummation is also discussed in Refs.\,\cite{Sterman:2000pt, Mukherjee:2006uu}.  That scheme involves a Mellin transform and Fourier transform, with the latter operation inviting two differing treatments: MFc, in which the cosine is retained, and MFe, where the cosine is approximated by unity because the argument introduces subleading corrections.  Ref.\,\cite{Aicher:2010cb} used MFc.

Ref.\,\cite{Barry:2021osv} adapted these approaches to produce independent fits whose large-$x$ behaviour is constrained by E615 data.  A fair representation of those $\zeta=\zeta_c$ fits is provided by the following parametrisation:
%%\begin{subequations}
%%\label{MFupizc}
%%\begin{align}
 %%&{\mathpzc u}_{\rm MF}^\pi(x;\zeta_c)  = n_{\mathpzc u} x^\alpha (1-x)^{\beta} (1+\gamma x^2)\,,\\
 %
 %%&\alpha = -0.15_{\mp 0.03}\,,\;
 %%\beta = 2.67_{\pm 0.07} \,,\;
 %%\gamma = 3.21_{\pm 0.58}  \,,  \label{E615JAMparamsMF}
%%\end{align}
%%\end{subequations}
%%
\begin{subequations}
\label{MFupizc}
\begin{align}
 &{\mathpzc u}_{\rm MF}^\pi(x;\zeta_c)  = n_{\mathpzc u} x^\alpha (1-x)^{\beta} (1+\gamma x^2)\,,\\
 &\alpha = -0.53_{\mp 0.03}\,,\;
 \beta = 2.41_{\pm 0.01} \,,\;
 \gamma = 16.5_{\pm 0.6}  \,,  \label{E615JAMparamsMF}
\end{align}
\end{subequations}
which represent an average of the MFc and MFe results and express an effective large-$x$ exponent $\beta(\zeta_c) = 2.45(11)$.

\begin{figure}[t]
\includegraphics[width=0.435\textwidth]{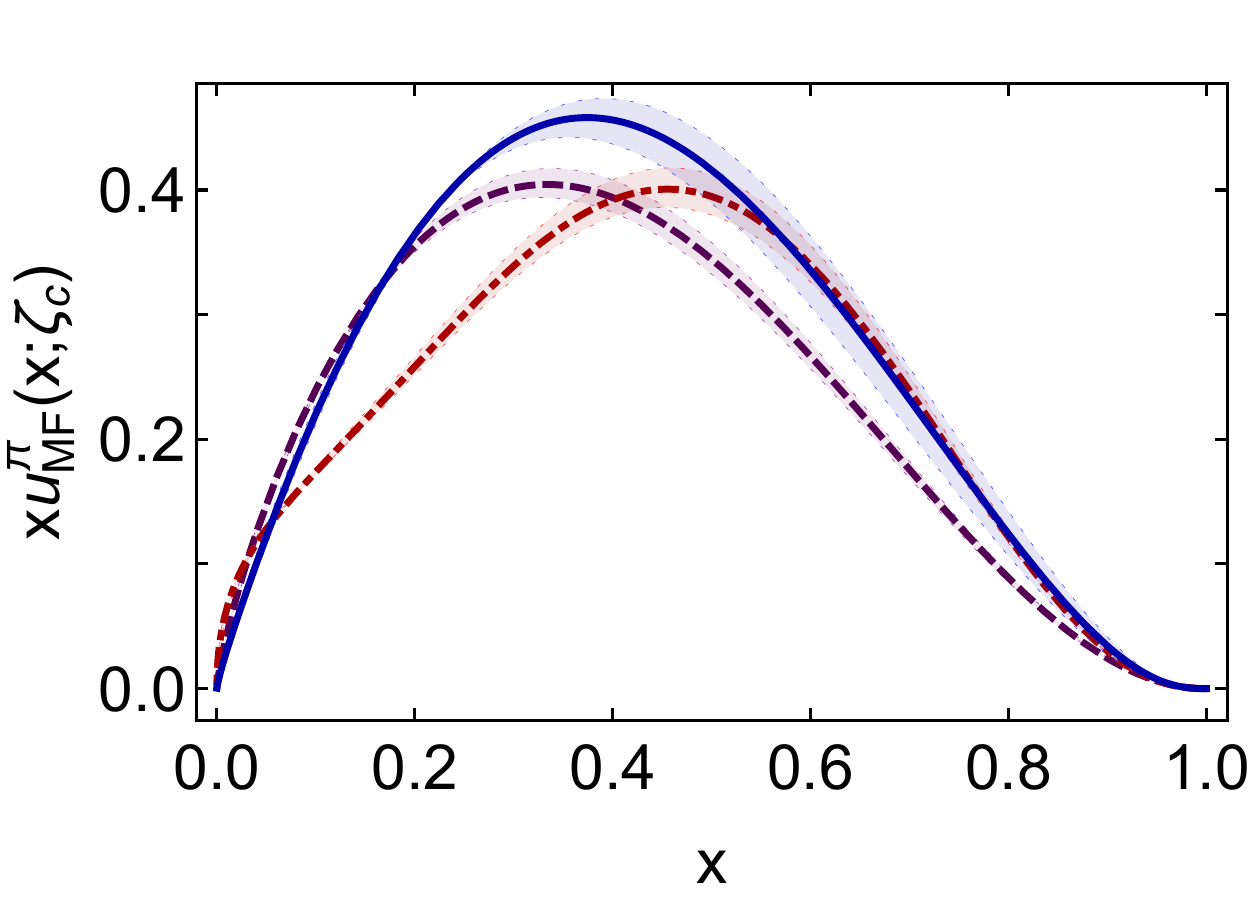}
\caption{\label{zetaccfDingCui}
Valence-quark DF at $\zeta=\zeta_c=1.27\,$GeV.
Dot-dashed red curve within like-coloured band -- Eqs.\,\eqref{MFupizc}, derived from Ref.\,\cite{Barry:2021osv}, with $\beta_{\rm eff}(\zeta_c) = 2.45(11)$;
solid blue curve and band -- prediction from Refs.\,\cite{Cui:2020dlm, Cui:2020tdf, Chang:2021utv}, which express $\beta_{\rm eff}(\zeta_c) = 2.52(5)$;
and dashed purple curve and band -- Eqs.\,\eqref{MFupizcS}, $\beta_{\rm eff}(\zeta_c) = 2.06(2)$.
}
\end{figure}

Matching the outcome of the dM analysis, the Ref.\,\cite{Barry:2021osv} MF fits represented by Eqs.\,\eqref{MFupizc} produce $\langle 2 x \rangle^{\zeta_c}_{{\mathpzc u}_{\rm MF}}=0.47(2)$, \emph{i.e}., lodge a markedly smaller fraction of the pion's light-front momentum with the valence quarks than existing calculations predict.
This is highlighted in Fig.\,\ref{zetaccfDingCui} by comparison with the parameter-free prediction of the pion valence-quark DF in Refs.\,\cite{Cui:2020dlm, Cui:2020tdf, Chang:2021utv}, evolved from the $\zeta=\zeta_{\cal H}$ result:
\begin{align}
{\mathpzc u}^\pi&(x;\zeta_H) = 375.32\,x^2 (1-x)^2 \nonumber \\
& \quad \times
[1-2.5088 \sqrt{x(1-x)} + 2.0250 x(1-x)]^2, \label{FpionPDF}
\end{align}
whose dilated (hardened) profile owes to EHM.  Plainly, the area under the dot-dashed red curve is less than that under the solid blue curve.

Fig.\,\ref{zetaccfDingCui} exposes a curious feature of the MF fits in Ref.\,\cite{Barry:2021osv}; namely, the DFs are not of uniform concavity on $x\lesssim 0.5$.  Potential causes of such ``wiggles'' in the Ref.\,\cite{Barry:2021osv} fits are
limitations introduced by the simple DF fitting \emph{Ansatz} employed, written in Eq.\,\eqref{MFupizc}, and/or
choosing to treat valence, glue, and sea DFs as uncorrelated at $\zeta_c$.
These particular features are not evident in the dM results, Eqs.\,\eqref{dMupizc}, possibly because DFs consistent with {\sf T1} have a rich structure, whose reliable extraction requires a more nuanced approach to fitting.

We have checked the DF in Eqs.\,\eqref{MFupizc} against the {\sf P1} corollary in Eq.\,\eqref{RecursionAnyz}, with the result displayed in Fig.\,\ref{FigOddMoments}.  Evidently, the recursion relation is not satisfied.  This failure can be traced to the wiggles on $x\lesssim 0.5$.  They can be eliminated as follows.
(\emph{i}) Using Eq.\,\eqref{EqxnzzH}, evolve the DF in Eq.\,\eqref{MFupizc} to $\zeta_{\cal H}$, whereat $\langle 2 x \rangle_{{\mathpzc u}_{\rm MF}^\pi}^{\zeta_{\cal H}} = 1$.
(\emph{ii}) Working with the first $101$ Mellin moments at $\zeta_{\cal H}$, determine the best least-squares fit achievable with a symmetric function.  The form
\begin{equation}
{\mathpzc u}^\pi(x) = n_{\mathpzc u} x^\alpha (1-x)^\alpha [1-\sqrt{x(1-x)}]^2
\end{equation}
is efficacious: $\alpha = 1.28(9)$ reproduces the moments with mean absolute relative error $6(5)$\%.
(\emph{iii}) Evolve the symmetric functions back to $\zeta_c$ using Eq.\,\eqref{EqxnzzH}.
This procedure delivers DFs that may be approximated as
\begin{subequations}
\label{MFupizcS}
\begin{align}
 \tilde {\mathpzc u}_{\rm MF}^\pi(x;\zeta_c)  & = n_{\mathpzc u} x^\alpha (1-x)^{\beta} (1- x^2)\,,\\
 &\alpha = -0.20_{\mp 0.03}\,,\;
 \beta = 1.09_{\pm 0.02}  \,,  \label{E615JAMparamsMFS}
\end{align}
\end{subequations}
yield $\langle 2 x \rangle_{\tilde{\mathpzc u}_{\rm MF}^\pi}^{\zeta_{c}} =0.465(12)$, express an effective large-$x$ exponent $\beta(\zeta_c) = 2.06(2)$, and are represented in Fig.\,\ref{zetaccfDingCui} by the dashed purple curve and linked band.
Evidently, requiring elimination of the wiggles forces the peak of $x {\mathpzc u}_{\rm MF}^\pi$ to shift toward $x=0$ whilst maintaining the area under the curve.

\begin{figure}[t]
\vspace*{2ex}

\leftline{\hspace*{0.5em}{\large{\textsf{A}}}}
\vspace*{-4.5ex}
\includegraphics[width=0.435\textwidth]{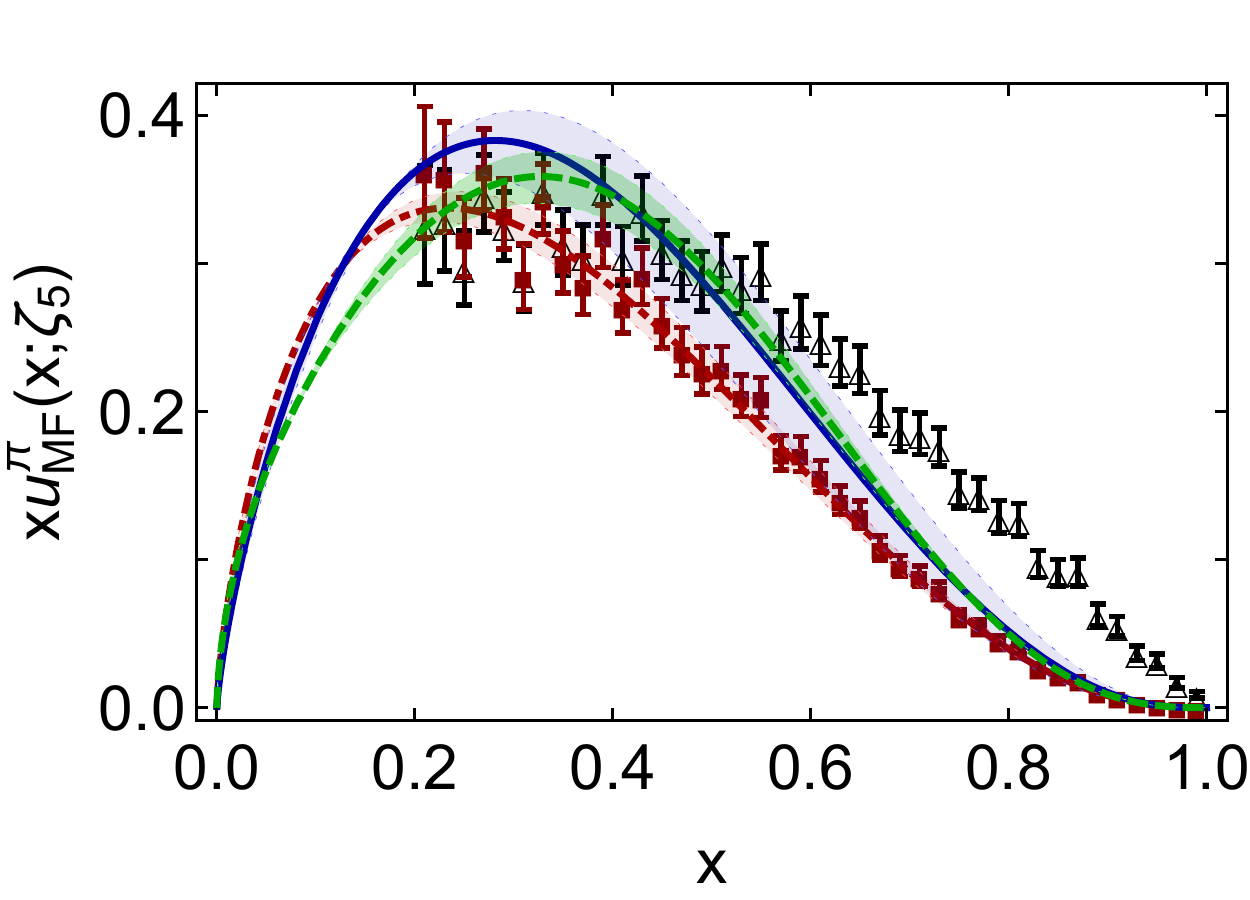}
\vspace*{-1ex}

\leftline{\hspace*{0.5em}{\large{\textsf{B}}}}
\vspace*{-3ex}
\includegraphics[width=0.435\textwidth]{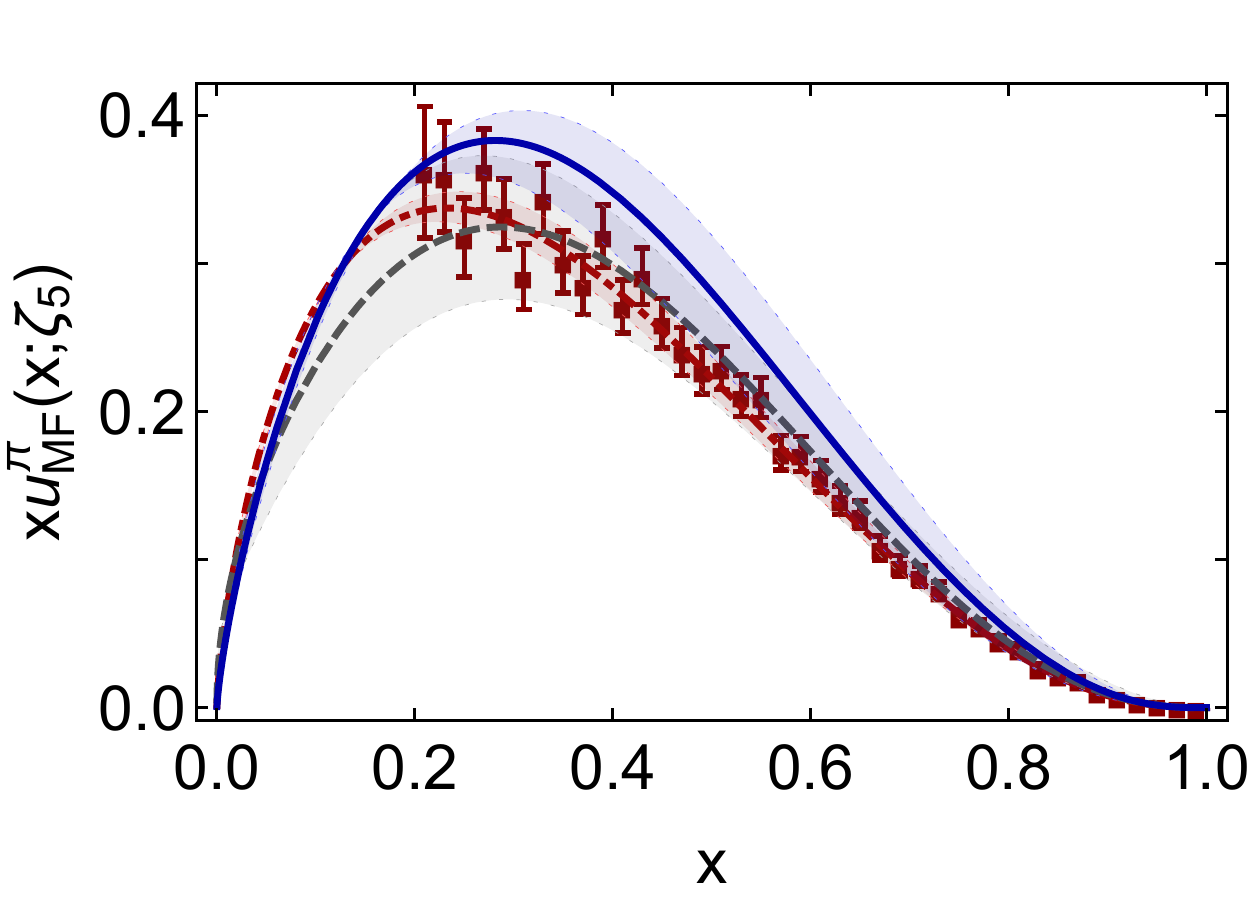}
\caption{\label{FMFE615}
\emph{Upper panel}\,--\,{\sf A}.
Pion valence-quark DF (dot-dashed red curve) inferred from fits to data in Ref.\,\cite{Barry:2021osv} obtained with NLL threshold resummation performed using the MF method, evolved $\zeta_c \to \zeta_5$:  $\langle 2 x \rangle^{\zeta_5}_{\tilde{\mathpzc u}_{\rm MF}}=0.36(1)$
[The encompassing band expresses the uncertainty in Eqs.\,\eqref{MFupizcS}.]
At this scale, the large-$x$ exponent is $\beta = 2.24(7)$.
Data recorded in Ref.\,\cite[E615]{Conway:1989fs} -- black open up-triangles;
and red squares -- projection of that data onto the $\zeta_c \to \zeta_5$ evolution of Eq.\,\eqref{MFupizcS}, which we denote as E615$_{\rm MF}$.
Parameter-free prediction from Refs.\,\cite{Cui:2020dlm, Cui:2020tdf, Chang:2021utv}: blue curve within like-coloured band, which yields $\langle 2 x \rangle^{\zeta_5}_{{\mathpzc u}}=0.41(4)$.
MFc fit to data \cite[E615]{Conway:1989fs} in Ref.\,\cite{Aicher:2010cb}: dashed green curve within like-coloured band, $\langle 2 x \rangle^{\zeta_5}_{{\mathpzc u}_{\rm MFc}}=0.40(2)$.
\emph{Lower panel}\,--\,{\sf B}.  As in {\sf A}, except: data from Ref.\,\cite[E615]{Conway:1989fs} and fit from Ref.\,\cite{Aicher:2010cb} are removed; and lattice-QCD prediction from Ref.\,\cite{Sufian:2019bol} is included, drawn as dashed grey curve within like-coloured band.
}
\end{figure}

The DFs approximated by Eqs.\,\eqref{MFupizcS} reproduce the first $101$ moments of the DFs in Eq.\,\eqref{MFupizc} with mean absolute relative error $7(6)$\%; and in being uniformly concave-down on $x\in (0,1)$, we judge their pointwise behaviour to be a more realistic portrayal of pion structure than Eqs.\,\eqref{MFupizc} -- modern calculations of ground-state pseudoscalar-meson LFWFs, e.g., Refs.\,\cite{Chang:2013pq, Li:2016dzv, Chouika:2017rzs, Ding:2018xwy, Qian:2020utg, Zhang:2020gaj, Choi:2020xsr, dePaula:2020qna, Lu:2021sgg, Raya:2021zrz} do not exhibit features that could produce wiggles in the associated valence-quark DFs.
Consequently, we hereafter represent Eqs.\,\eqref{MFupizcS} as a fair and physical sketch of the MF results in Ref.\,\cite{Barry:2021osv}.
Expressed another way, we interpret Eqs.\,\eqref{MFupizcS} as a smoothed approximation to Eqs.\,\eqref{MFupizc}, equivalent in (most) practically measurable respects.
It is conceivable that Eqs.\,\eqref{MFupizcS} could emerge following refinements of the analysis\linebreak scheme in Ref.\,\cite{Barry:2021osv}; but given that {\sf C1} applies to Eqs.\,\eqref{MFupizcS}, that would require introduction of intimate correlations between the valence, glue, and sea DFs.

Evolved to $\zeta=\zeta_5=5.2\,$GeV, the DFs associated with Eqs.\,\eqref{MFupizcS} are drawn in Fig.\,\ref{FMFE615}A.  The pion valence-quark DF inferred in Ref.\,\cite{Aicher:2010cb} from E615 data is also drawn in Fig.\,\ref{FMFE615}A: on $x\gtrsim 0.3$ it agrees with the prediction in Refs.\,\cite{Cui:2020dlm, Cui:2020tdf, Chang:2021utv} and yields a compatible valence-quark light-front momentum fraction.  The treatment of glue and sea distributions in Ref.\,\cite{Aicher:2010cb} is very different from that in Refs.\,\cite{Cui:2020dlm, Cui:2020tdf, Chang:2021utv}, explaining the $x\lesssim 0.3$ discrepancy.

\begin{figure}[t]
\vspace*{2ex}

\leftline{\hspace*{0.5em}{\large{\textsf{A}}}}
\vspace*{-4.5ex}
\includegraphics[width=0.435\textwidth]{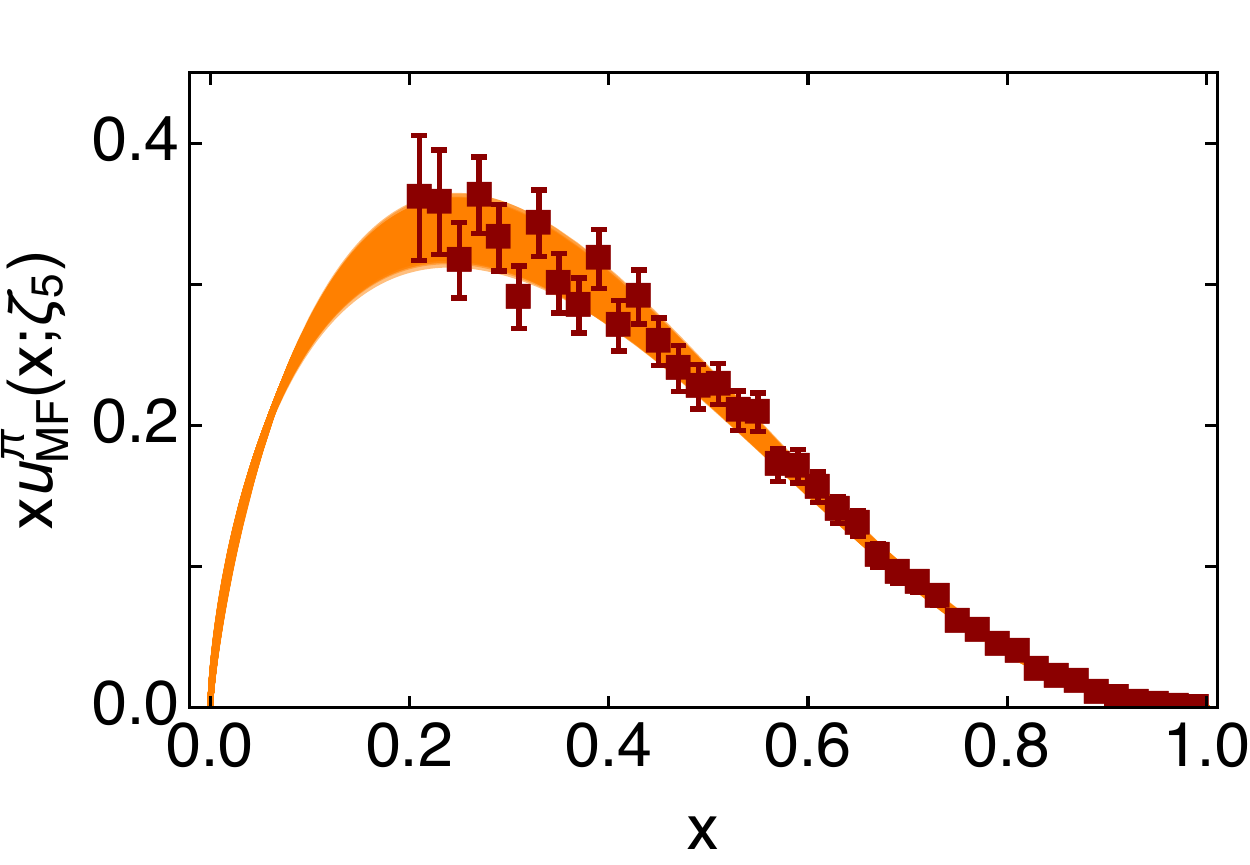}
\vspace*{+1ex}

\leftline{\hspace*{0.5em}{\large{\textsf{B}}}}
\vspace*{-3ex}
\includegraphics[width=0.435\textwidth]{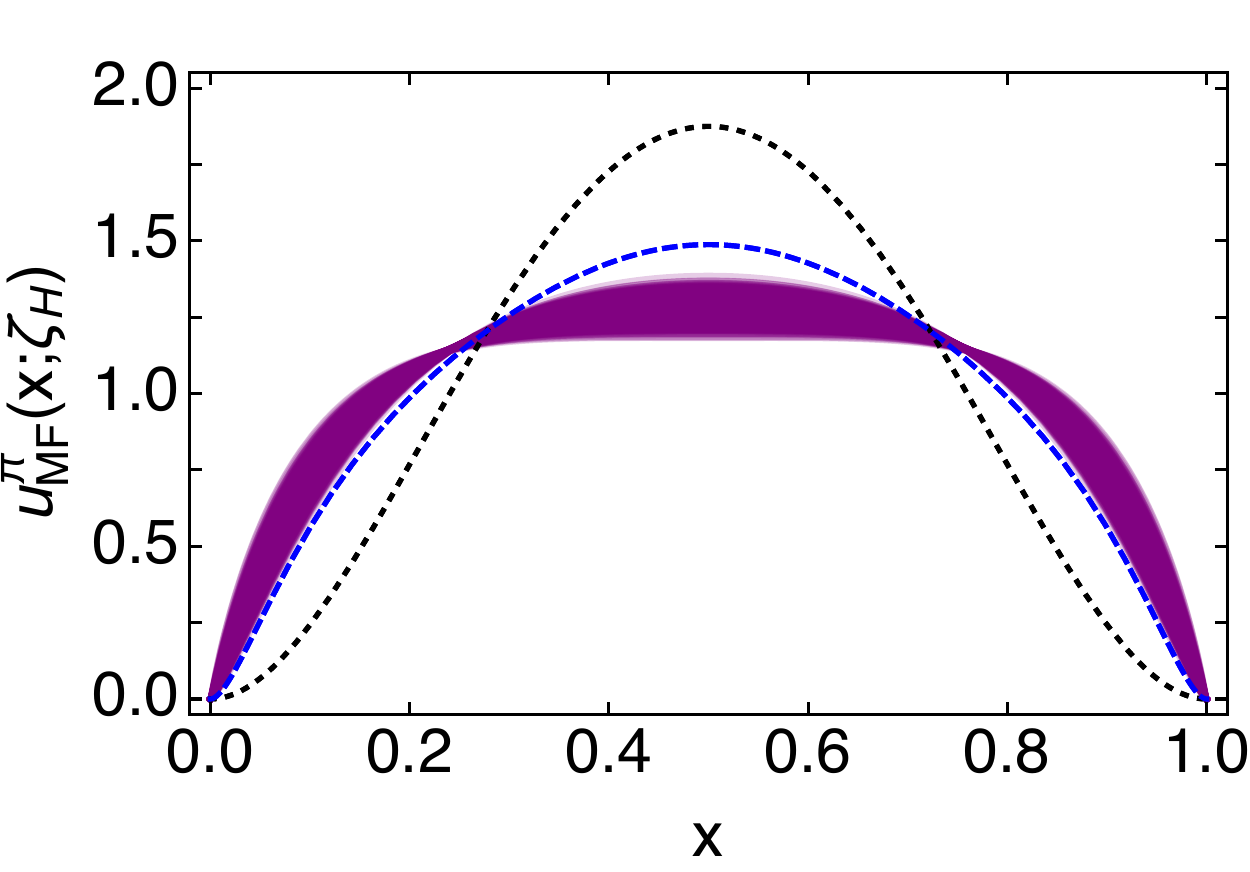}
\caption{\label{FMFE615K}
\emph{Upper panel}\,--\,{\sf A}.  Randomly distributed ensemble of pion valence-quark DFs (orange curves) constructed from the E615$_{\rm MF}$ data (red squares) using the procedure described in connection with Eq.\,\eqref{EqProb}.
\emph{Lower panel}\,--\,{\sf B}.  $\zeta_5\to \zeta_{\cal H}$ evolution of each curve in Panel {\sf A} (purple curves).  Parameter-free $\zeta_{\cal H}$ prediction in Refs.\,\cite{Cui:2020dlm, Cui:2020tdf, Chang:2021utv}, which expresses EHM-induced dilation: dashed blue curve.  Scale free DF, ${\mathpzc u}^{\rm sf}(x)=30 x^2(1-x)^2$: dotted black curve.
}
\end{figure}

Further to these points, working with the E615$_{\rm MF}$ data in Fig.\,\ref{FMFE615}, we repeated the analysis described in Sec.\,7.  The $K\gtrsim 1000$ replicas of this data are drawn in Fig.\,\ref{FMFE615K}A and their $\zeta_5\to \zeta_{\cal H}$ evolutions in Fig.\,\ref{FMFE615K}B: all evolved curves are consistent with {\sf C1}.
Moreover, comparison with the continuum prediction from Refs.\,\cite{Cui:2020dlm, Cui:2020tdf, Chang:2021utv}, Eq.\,\eqref{FpionPDF}, whose dilated (hardened) profile owes to EHM, demonstrates that, within mutual uncertainties, the set of evolved curves abuts the class of DFs linked to a quark+antiquark interaction with the $1/k^2$ ultraviolet behaviour which characterises QCD.\footnote{Analogous to Fig.\,\ref{FdME615K}B, the binning issue with the original E615 data is evident here in an overpopulation of evolved replicas with enhanced endpoint magnitudes.}
Hence, in contrast to the dM scheme, the MF approaches to NLL resummation in the analysis of Drell-Yan data lead to valence-quark DFs whose large-$x$ behaviour is approximately consistent with Eq.\,\eqref{pionDFpQCD} and can thus be connected with pion LFWFs whose momentum dependence matches QCD's prediction, Eq.\,\eqref{OPE}.  In this case, the data may be identified as resulting from measurements which are sensitive to intrinsic properties of the pion.

\begin{figure}[t]
\vspace*{2ex}

\includegraphics[width=0.435\textwidth]{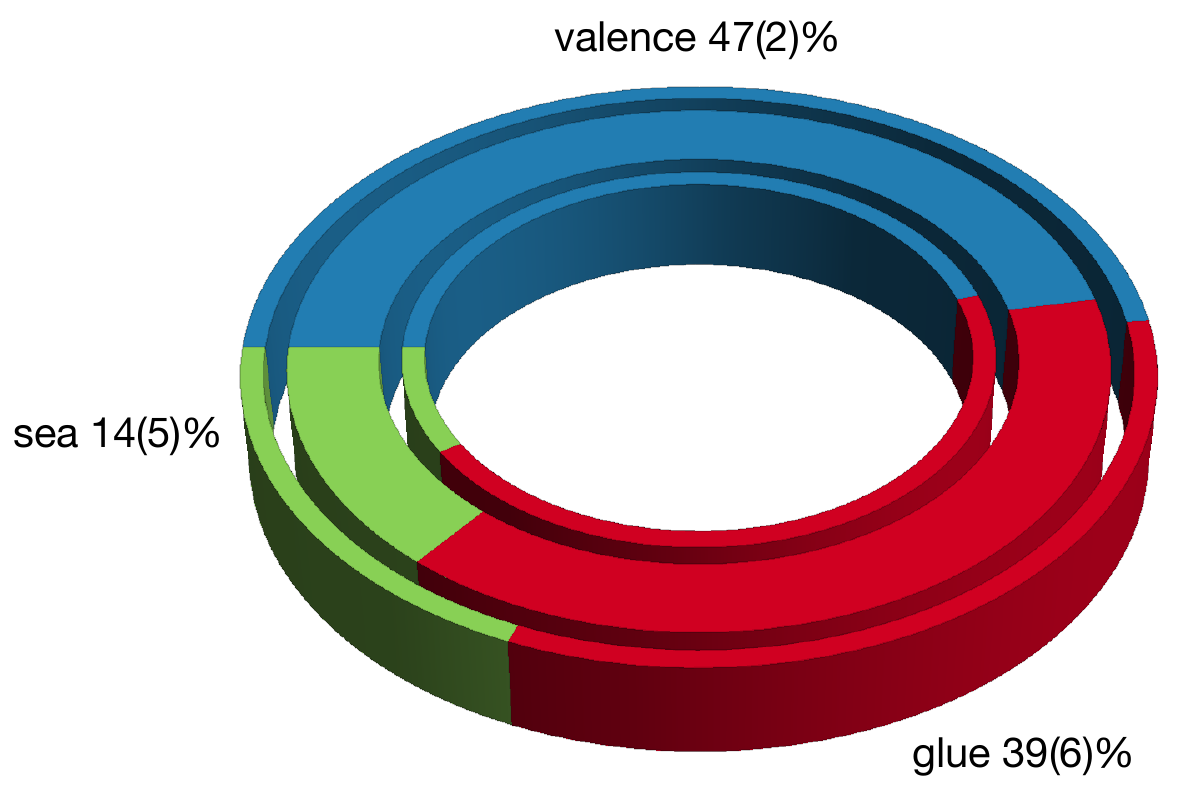}

\vspace*{-40.4ex}
\leftline{\hspace*{0.5em}{\large{\textsf{A}}}}

\vspace*{40ex}

\includegraphics[width=0.437\textwidth]{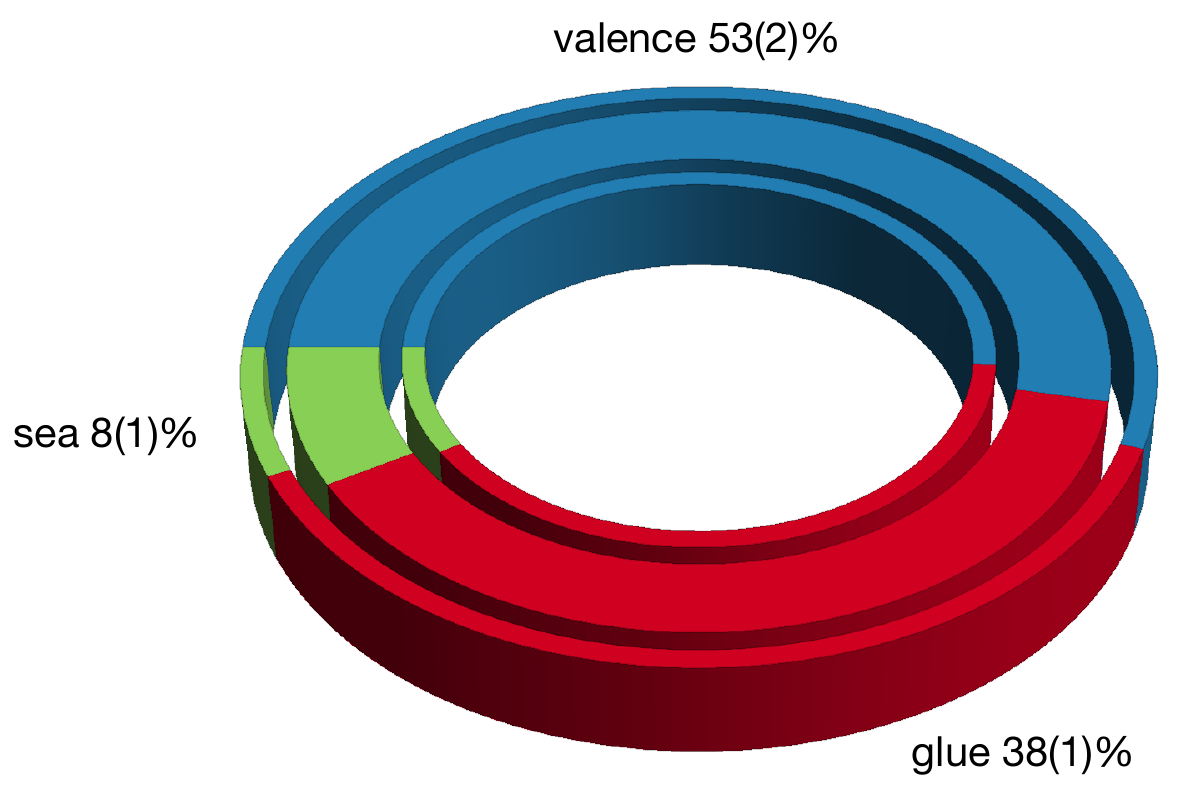}

\vspace*{-40.4ex}
\leftline{\hspace*{0.5em}{\large{\textsf{B}}}}

\vspace*{40ex}

\caption{\label{masssquared}
Pion mass-squared fractions carried by different parton species at $\zeta=\zeta_c=1.27\,$GeV.
\emph{Upper panel}\,--\,{\sf A}.  Row~2 in Table~\ref{TabMass2Fraction}, inferred in Ref.\,\cite{Barry:2021osv}
\emph{Lower panel}\,--\,{\sf B}.  Row~3 in Table~\ref{TabMass2Fraction}, calculated using the DF predictions in Refs.\,\cite{Cui:2020dlm, Cui:2020tdf, Chang:2021utv}.
}
\end{figure}

With the information now at hand, it is possible to follow Ref.\,\cite[Sec.\,VIII.A]{Raya:2021zrz} and record the pion mass-squared fractions carried by the different parton species at $\zeta=\zeta_c$.  The results are listed in Table~\ref{TabMass2Fraction} and illustrated in Fig.\,\ref{masssquared}.  Evidently, despite the marked differences between their pointwise behaviour, the dM and MF fits in Ref.\,\cite{Barry:2021osv} yield practically the same pion mass-squared apportionments; and compared with the predictions associated with Eq.\,\eqref{FpionPDF}, they lodge far more of $m_\pi^2$ with the sea-quarks at the cost of the valence-quark fraction.
Curiously, the relative stability of the valence-quark fraction indicates that the sea fraction increases as the glue fraction decreases and vice versa.  This is counterintuitive, given the character of QCD evolution, with gluon splitting serving to populate the sea.  Hence, it is likely an artefact in Ref.\,\cite{Barry:2021osv} of treating parton DFs as independent at $\zeta_c$ in the fitting.

\begin{table}[t!]
\caption{%\small
\label{TabMass2Fraction}
Pion mass-squared fractions carried by different parton species at $\zeta=\zeta_c=1.27\,$GeV.
}
\begin{tabular*}%{|c|c|c|c|c|c|c|}\hline
{\hsize}
{
l@{\extracolsep{0ptplus1fil}}
|l@{\extracolsep{0ptplus1fil}}
l@{\extracolsep{0ptplus1fil}}
l@{\extracolsep{0ptplus1fil}}}\hline
%%\begin{tabular}{l|c|c|c|c|c|c|c}\hline
source & {\rm valence} & {\rm glue} & {\rm sea} \\\hline
\mbox{\cite{Barry:2021osv}}$_{\rm dM}\ $ & 0.46(3) & 0.40(5) & 0.15(7) \\
%% 0.46(3) 0.15(7) 0.40(5)
\mbox{\cite{Barry:2021osv}}$_{\rm MF}\ $ & 0.47(2) & 0.39(6) & 0.14(5) \\\hline
\mbox{\cite{Cui:2020tdf}}$\ $ & 0.53(2) & 0.38(1) & 0.084(9)\\\hline
\end{tabular*}
\end{table}

%\medskip

%\noindent\emph{9.$\;$Summary and perspectives} ---
\section{Summary and outlook}
\label{epilogue}
QCD predictions for the behaviour of the pion Bethe-Salpeter- and light-front wave functions at large relative momenta have been known for more than forty years; and we have highlighted [Sec.\,\ref{SecLargeX}] that all calculations of the pion valence-quark distribution function (DF) which are consistent with these predictions deliver the following result:
\begin{equation}
\label{pionDFpQCDFinal}
{\mathpzc u}^\pi(x;\zeta) \stackrel{x\simeq 1}{\sim} (1-x)^{\beta \,=\,2+\gamma(\zeta)}\,,
\end{equation}
where $\gamma(\zeta) \geq 0$ grows logarithmically with $\zeta$.  Equivalently, using any known evolution scheme, no description of pion structure that is consistent with QCD pion wave functions can produce a large-$x$ exponent $\beta<2$ on $\zeta \gtrsim m_p$, where $m_p$ is the proton mass.

Yet some approaches to the analysis of extant data, which should serve to constrain the large-$x$ behaviour of ${\mathpzc u}^\pi(x;\zeta>m_p)$, nevertheless violate Eq.\,\eqref{pionDFpQCDFinal}.
For instance, whereas fits based on next-to-leading-order perturbative QCD treatments of hard scattering kernels that include soft gluon resummation using the Mellin-Fourier (MF) scheme yield DFs in agreement with\linebreak Eq.\,\eqref{pionDFpQCDFinal}, those which use the double-Mellin (dM) approach to soft gluon resummation do not.
This does not necessarily mean that the MF approaches are the more rigorous, but it does justify continued efforts aimed at understanding their efficacy in delivering DFs that are consistent with QCD predictions.
On the other hand, if one chooses to favour the dM approach, for some reason or another, then the associated disagreement with Eq.\,\eqref{pionDFpQCDFinal} requires explanation; and these are the only possibilities: [\emph{a}] the dM scheme is incomplete, omitting or misrepresenting some aspect or aspects of the hard processes involved; [\emph{b}] (some of) the data being considered in the analysis are not a true expression of a quality intrinsic to the pion; or [\emph{c}] QCD, as it is currently understood, is not the theory of strong interactions.

In supporting these conclusions, we explained an array of corollaries that follow from a single proposition [{\sf P1} in Sec.\,\ref{SecAOE}]: there exists an effective charge that defines an evolution scheme for parton DFs that is all-orders exact.  A great deal can be concluded from this proposition without specifying the form of the charge.
For instance, {\sf P1} entails
that there is a scale $\zeta =\zeta_{\cal H} \in [0,m_p)$ whereat valence-quarks carry all the pion's light-front longitudinal momentum; $\gamma(\zeta_{\cal H})\equiv 0$; and, consequently, glue and sea distributions are
completely determined by the nonperturbative information embedded in ${\mathpzc u}^\pi(x;\zeta_{\cal H})$ and revealed in evolution from this scale.
%completely determined by evolution from the nonperturbative information embedded in ${\mathpzc u}^\pi(x;\zeta_{\cal H})$.
%
These outcomes are guaranteed when ${\mathpzc u}^\pi(x;\zeta_{\cal H})={\mathpzc u}^\pi(1-x;\zeta_{\cal H})$, a quality expressed in all calculations that respect Poincar\'e covariance and QCD's vector and axial-vector Ward-Green-Takahashi identities.  Such symmetry entails that any given odd Mellin moment of ${\mathpzc u}^\pi(x;\zeta)$ is linearly dependent upon the even Mellin moments of lower order.  This corollary is expressed in a recursion relation that can be used to reveal the character of any DF, whether fitted or calculated.

Forty years after the first experiment to collect data amenable for use in constraining the large-$x$ behaviour of ${\mathpzc u}^\pi(x)$, the answer, for some practitioners, remains uncertain.  In large part, the paucity of such data and its imprecision are responsible.  Modern and anticipated facilities promise to remedy these issues.  New developments in phenomenology and theory are required to eliminate the others.

%
%\section*{Acknowledgments}
\medskip
\noindent\emph{Acknowledgments}.
We are grateful for constructive comments from K.-L.~Cai, O.~Denisov, F.~de~Soto, T.~Frederico, J.~Friedrich, C.~Mezrag, V.~Mokeev, W.-D.~Nowak, C.~Quintans, G.~Salm\`e and J.~Segovia.
Work supported by:
National Natural Science Foundation of China (grant nos.\,12135007, 11805097);
Helmholtz-Zentrum\linebreak Dresden-Rossendorf High Potential Programme;
Spanish Ministry of Science and Innovation (MICINN) (grant nos.\ PID2020-113334GB, PID2019-107844GB-C22);
Generalitat Valenciana (grant no.\ Prometeo/2019/087);
Junta de Andaluc{\'{\i}}a (grant nos.\ P18-FR-5057, UHU-1264517, UHU EPIT-2021);
and STRONG-2020 ``The strong interaction at the frontier of knowledge: fundamental research and applications'' which received funding from the European Union's Horizon 2020 research and innovation programme (grant no.\,824093).

%\bibliographystyle{../../../zProc/z10/z10KITPC/h-physrev4}
%%\bibliographystyle{elsarticle-num-names}
%%\bibliography{../../../../CollectedBiB}

\end{document}